\documentclass[preprint,authoryear, 12pt]{elsarticle}
\usepackage{amsmath,amssymb,mathrsfs}

\usepackage{graphicx}
\usepackage{subfigure}

\usepackage{mathtools}
\usepackage{amsfonts}
\usepackage{cases}
\usepackage{color}
\usepackage{natbib}
\usepackage{bbm}
\usepackage{enumerate}

\usepackage[thmmarks]{ntheorem}
{
\theoremstyle{nonumberplain}
\theoremheaderfont{\bfseries}
\theorembodyfont{\normalfont}
\theoremsymbol{\mbox{$\Box$}}

}
\newtheorem{thm}{Theorem}[section]
\newtheorem{lem}{Lemma}[section]

\newtheorem{rem}{Remark}[section]
\newtheorem{pr}{Proposition}[section]

\newtheorem{cor}{Corollary}[section]
\newtheorem{asp}{Assumption}[section]

\usepackage{titlesec}

\usepackage{geometry}
\journal{arXiv.org}

\begin{document}
\begin{frontmatter}
\title{Optimal consumption and portfolio selection with Epstein-Zin utility under general constraints}
\author[math]{Zixin Feng}
\ead{FengZiXinFZX@163.com}
\author[math]{Dejian Tian\corref{correspondingauthor}}
\ead{djtian@cumt.edu.cn}
\address[math]{School of Mathematics, China University of Mining and Technology, Xuzhou, P.R. China}
\cortext[correspondingauthor]{Corresponding author}

\begin{abstract}
The paper investigates the consumption-investment problem for an investor with Epstein-Zin utility  in an incomplete market.  Closed, not necessarily convex, constraints  are imposed on strategies.  The optimal consumption and investment strategies are characterized via a quadratic backward stochastic differential equation (BSDE).  Due to the stochastic market environment, solutions to this BSDE are unbounded and thereby the BMO argument breaks down.  After establishing the martingale optimality criterion,  by delicately selecting Lyapunov functions,  the verification theorem is ultimately obtained.  Besides,  several examples  and  numerical simulations for the optimal strategies are provided and illustrated.
\end{abstract}

\begin{keyword}
Epstein-Zin utility; quadratic BSDE; consumption-investment problem; closed constraints.
\end{keyword}

\end{frontmatter}

\section{Introduction}
In the classical representative agent framework, the investor's preferences are mostly characterized by time-separable utility functions.  Optimal consumption-investment problem for this kind of utility  has been developed comprehensively by numerous researchers. Originally articulated in the context of the Markovian structure in the landmark paper by \cite{M71}, the theory was later extended to non-Markovian models by \cite{P86}, \cite{KLS87} and \cite{CH89} using the martingale method in a complete market.  \cite{KS99} gave a duality result and \cite{HIM05} investigated the problem by the method of backward stochastic differential equations (BSDEs) for a class of time separable utilities in an incomplete market.  See \cite{P08} for more related problems with time separable utilities.

However, widely used time separable utilities unintentionally impose an artificial relation between risk aversion $\gamma$ and elasticity of intertemporal substitution (EIS) $\psi$: they are reciprocal to each other. Such a relationship will lead to a large number of asset pricing anomalies, such as the low risk premium and high risk-free rate. To disentangle risk aversion from EIS,  the notion of recursive utilities was first specified in discrete time by \cite{EZ89}. Then its continuous-time analog was formulated in \cite{DE92}. These Epstein-Zin type utilities provide a framework to tackle the aforementioned asset puzzles.  The readers can refer to \cite{B07, BY04,BCG11} for more explanations and clarifications on $\gamma$ and $\psi$.

Over the past two decades, there has been substantial progress for the consumption and investment problem with recursive utility of Epstein-Zin type in stochastic market environments.  Using  the utility gradient approach, \cite{SS99} studied the optimal strategies for Epstein-Zin utility with parameter $\theta=\frac{1-\gamma}{1-1/\psi}<0$.  \cite{KSS13} and  \cite{KSS17} investigated the problem for Epstein-Zin utility with assumption (H) on parameters excluding $\gamma>1$ and $\psi>1$ by the tool of Hamilton-Jacobi-Bellman (HJB) equations.  Motivated by empirical evidences and observations suggesting the parameters $\gamma>1$ and  $\psi>1$, \cite{X17} considered the corresponding problem for Epstein-Zin utility with the advantage of BSDE techniques.  Besides, \cite{MX18} introduced a dual approach to study the consumption-investment problem for Epstein-Zin utility with parameters  $\gamma\psi>1$ and  $\psi>1$.

In the vast majority of the literature, it is often assumed that the investor is able to select his consumption and portfolio strategies with some constraints. For time-additive utilities, \cite{CK92} studied the stochastic control problem of maximizing expected utility from terminal wealth when the portfolio is constrained to take values in a closed convex set. One can refer to \cite{RE00} and \cite{BCX19} for more information about convex trading constraints. More importantly, \cite{HIM05} solved the optimal investment problem with \textit{closed but not necessarily convex set constraints} for time-additive utilities, such as exponential utility or constant relative risk aversion (CRRA) utility.  Later, \cite{CH11} introduced closed constraints for the consumption process based on the work of \cite{HIM05}.

For recursive utilities, \cite{EPQ01} stated a dynamic maximum principle to examine the consumption-investment problem with recursive utilities in the presence of nonlinear constraints on the wealth. \cite{SS03, SS05}  considered convex constraints on strategies.  Wang, Wang and Yang \cite{WWY16} developed a tractable incomplete market model with an earning process subject to permanent shocks and borrowing constraints. The readers can also refer to  \cite{AH19} and  \cite{MMS20} for maximizing Epstein-Zin utility with random horizons and transaction costs  respectively.

Motivated by \cite{HIM05} and \cite{CH11},  the current paper is committed to studying the optimal consumption and investment problem for Epstein-Zin utility with \textit{closed constraints on strategies} in an incomplete market whose parameters are driven by a state variable.  Using the same framework as in \cite{X17},  we focus on the specification $\gamma>1$ and $\psi>1$.   Compared to existing results, this paper contributes to the literature in the following three aspects.

First, we impose the constraints on strategies for Epstein-Zin utility, which is supposed to simply be closed.  To our best knowledge,  this general constraints have not been investigated for Epstein-Zin utility before.  Due to closed constraints on strategies,  the utility gradient approach and the dual method are no longer available.  Our method is based on BSDE techniques coming from \cite{HIM05} and \cite{X17}.  Comparing with \cite{X17},  the BSDE derived by the martingale optimal principle is more complicated, which is involved a projection term on a closed set.  Fortunately, this term does not essentially change the quadratic structure of the BSDE. However,  the distance function not only contains term $z$, but also involves the unbounded market parameters. This fact presents difficulties in proving the properties of BSDE's solutions.  After careful estimations, the upper boundedness of the solution still holds under appropriate assumptions.  It deserves pointing out that we present the martingale optimal principle (Theorem \ref{g}) as one of our main results, which is implicit in \cite{X17}.

Second,  our model admits market parameters unbounded. Solutions to this quadratic BSDE are unbounded and thereby the often-used BMO argument breaks down. Closed constraints on the set of admissible strategies result in a situation that  the candidate optimal strategy cannot be explicitly expressed by $Z$, one part of solutions to the BSDE.  To prove this exponential local martingale (induced by the stochastic integral term of the optimal strategy) is a martingale,  we need to carry out more sophisticated estimations and choose delicately a suitable Lyapunov function to overcome the difficulties. Unlike the situation of \cite{X17}, we find  the effect of constraints on portfolio causes the Lyapunov argument may fail for sufficiently large $\gamma$. We propose the condition $\frac{1}{2}+\frac{2(1-\gamma)}{\gamma}\left|\sigma'\Sigma^{-1}\sigma\rho\right|\left|\rho\right|>0$ to proceed the argument, and illustrate the universality of the method.  Consequently, the verification theorem holds.

Third, we provide three specific numerical examples to illustrate the influence of the constraints and comparative statics analysis on the optimal strategy.  The first one is Black-Scholes model.  As far as we know, Epstein-Zin utility model with closed constraints has not been investigated before even in this simplest market model. We change the portfolio constraints to show its impact on the portfolio and consumption.  Linear diffusion model and Heston stochastic volatility model are also investigated.  The former model corresponds to a bounded risk premium and volatility,  while in the latter case both risk premium and volatility are unbounded. We find the constraints have a smaller impact on consumption, but a larger impact on investment portfolios.

The remainder of this paper is organized as follows. Section 2 introduces the Epstein-Zin utility process.  Section 3 presents the stochastic market environment,  the consumption-investment problem and main results including the martingale optimal principle and the verification theorem.  Several examples  and  numerical simulations for the optimal strategies are provided and illustrated in Section 4.  All the proofs are relegated to Section 5.  Section 6 concludes the paper.

\section{Epstein-Zin preferences}

Given a time horizon $T<+\infty$. Let $\left(\Omega,\mathscr{F},\left(\mathscr{F}_t\right)_{0\leq t\leq T},\mathbb{P}\right)$ be a filtered probability space. $\left(\mathscr{F}_t\right)_{0\leq t\leq T}$ is the natural filtration generated by a $(k+n)$-dimensional standard Brownian motion $B=\left(W,W^\perp\right)$, where $W$ and $W^\perp$ are the first $k$ and the last $n$ components. We also assume $\left(\mathscr{F}_t\right)_{0\leq t\leq T}$ satisfies the usual hypotheses, completeness and right-continuity.

Let $\mathcal{R}_+$ be the set of all nonnegative progressively measurable processes on $[0,T]$. For $c\in\mathcal{R}_+$, if $t<T$, $c_t$ represents the consumption rate at time $t$; $c_T$ stands for a lump sum consumption at $T$. Throughout the whole paper, we always assume $\delta>0,\gamma>1$ and $\psi>1$, which stand for the discounting rate, the relative risk aversion and the EIS respectively. Given the bequest utility function $U(c)=\frac{c^{1-\gamma}}{1-\gamma}$, the Epstein-Zin utility for the consumption stream $c\in\mathcal{R}_+$ over a time horizon $T$ is a semimartingale $V^c$ that satisfies
\begin{align}\label{EZuc}
	V_t^c=\mathbf{E}_t\left[\int_t^T f(c_s, V_s^c)ds+U(c_T)\right],~~~\forall t\in[0,T],
\end{align}
where $\mathbf{E}_t(\cdot)$ denotes $\mathbf{E}(\cdot|\mathscr{F}_t)$ throughout the paper and $f:[0,+\infty)\times(-\infty,0]\to\mathbb{R}$ stands for the Epstein-Zin aggregator denoted by
\begin{align}\label{EZa}
f(c,v)&=\delta\frac{(1-\gamma)v}{1-\frac{1}{\psi}}\left[\left(\frac{c}{((1-\gamma)v)^{\frac{1}{1-\gamma}}}\right)^{1-\frac{1}{\psi}}-1\right]\notag\\
&=\frac{\delta c^{1-\frac{1}{\psi}}}{1-\frac{1}{\psi}}\left((1-\gamma)v\right)^{1-\frac{1}{\theta}}-\delta\theta v,
\end{align}
where $$\theta:=\frac{1-\gamma}{1-\frac{1}{\psi}}<0.$$
By Proposition 2.2 in \cite{X17},  $V^c$ can be characterized by the following BSDE
\begin{align}\label{EZub}
	V_t^c=U(c_T)+\int_t^Tf(c_s,V_s^c)ds-\int_t^TZ_s^cdB_s,~~~t\in[0,T].
\end{align}
Specifically, let $\mathcal{C}_a$ denote the class of consumption streams
$$\mathcal{C}_a:=\left\{c\in\mathcal{R}_+:\mathbf{E}\left[\int_0^{T}e^{-\delta s}c_s^{1-\frac{1}{\psi}}ds\right]<+\infty, \mathbf{E}\left[c_T^{1-\gamma}\right]<+\infty\right\}.$$
Then for each $c\in\mathcal{C}_a$, BSDE (\ref{EZub}) admits a unique solution $(V^c,Z^c)$ in which $V^c$ is continuous, strictly negative, of class D, and $\int_0^T|Z_t^c|^2dt<+\infty$, a.s.

\begin{rem}
It is worth noting that BSDE (\ref{EZub}) only requires the integrable terminal condition. The readers can refer to Theorem 1 of \cite{F19} for general results.
\end{rem}
\begin{rem}
Among the literature on Epstein-Zin utility, there exists another kind of integrability requirement on consumption streams
$$\mathbf{E}\left[\int_0^Tc_t^ldt+c_T^l\right]<+\infty,\quad\forall l\in\mathbb{R}.$$
One can refer to \cite{SS99,SS03,SS05}, \cite{KSS13} and \cite{KSS17}. However, these conditions are not general enough to capture all relevant consumption plans in our applications. $\mathcal{C}_a$ is the same as the one considered in \cite{X17}, which is weaker than aforementioned conditions.
\end{rem}

\section{The consumption-investment problem}
Motivated by \cite{HIM05}, \cite{CH11} and  \cite{X17}, we will investigate the consumption-investment optimization with Epstein-Zin utility under general constraints in an incomplete market.  Following the financial market framework of \cite{X17}, we introduce closed constraints on consumption and investment strategies, and derive the martingale optimal principle (Theorem \ref{g}), from which the candidate optimal strategy can be deduced. Under some mild restrictions on market parameters, the verification theorem (Theorem \ref{main}) is given by the Lyapunov function argument.

\subsection{The model  setup}
Let $E$ be an open domain in $\mathbb{R}^k$, and define an $E$-valued state process
\begin{align}\label{state}
dX_t=b(t,X_t)dt+a(t,X_t)dW_t,\quad X_0=x\in E,
\end{align}
where $b, a:\mathbb{R}_+\times E\to\mathbb{R}^k$ are given Borel measurable functions. Consider a financial market model consisting of a riskfree asset $S^0$ and risky assets $S=(S^1,\cdots, S^n)$, which satisfy the dynamics
\begin{align*}
&dS_t^0=r(t,X_t)S_t^0dt,\\
&dS_t=\operatorname{diag}(S_t)\left[\left(r(t,X_t)\mathbf{1}_n+\mu(t,X_t)\right)dt+\sigma(t,X_t)dW_t^\rho\right],
\end{align*}
where diag($S$) is a diagonal matrix with the elements of $S$ on the diagonal. Here $\mathbf{1}_n$ is an $n$-dimensional vector with each entry 1. $W^\rho:=\int_0^\cdot\rho(s,X_s)dW_s+\int_0^\cdot\rho^\perp(s,X_s)dW^\perp_s$ is an $n$-dimensional Brownian motion with correlation functions $\rho:\mathbb{R}_+\times E\to\mathbb{R}^{n\times k}$ and $\rho^\perp:\mathbb{R}_+\times E\to\mathbb{R}^{n\times n}$ satisfying $\rho\rho'+\rho^\perp\left(\rho^\perp\right)'=I_n$. Model coefficients $r: \mathbb{R}_+\times E\to\mathbb{R},\mu:\mathbb{R}_+\times E\to\mathbb{R}^n,\sigma:\mathbb{R}_+\times E\to\mathbb{R}^{n\times n}$ are all given Borel measurable functions.

Let $\mathcal{R}^n_+$ be the set of all predictable processes taking their values in $\mathbb{R}^n$. We allow constraints both on the investment strategy $\pi$ and the consumption process $c$. To this end, we introduce nonempty subsets $\Lambda\subseteq\mathcal{R}^n_+$ and $\mathcal{C}\subseteq\mathcal{C}_a$. Then we use the following concepts from Definition 3.1 and Definition 2.1 of \cite{CKV15}: $\Lambda$ (resp. $\mathcal{C}$) is \emph{sequentially closed} if it contains each process $\pi$ (resp. $c$) that is the $\lambda\otimes\mathbb{P}\raisebox{0mm}{-}a.s.$ limit of a sequence $(\pi^n)_{n\geq1}$ (resp. $(c^n)_{n\geq1}$) of processes in $\Lambda$ (resp. $\mathcal{C}$). $\Lambda$ (resp. $\mathcal{C}$) is $\mathcal{R}_+$-\emph{stable} if it contains $1_Ba+1_{B^c}a'$ for all $a, a'\in \Lambda$ (resp. $a, a'\in\mathcal{C}$) and each predictable set $B\subseteq[0,T]\times\Omega$. See more details in \cite{CH11}.

Suppose that $\Lambda$ and $\mathcal{C}$ satisfy the following assumption.
\begin{asp}\label{ccc}
$\Lambda$ and $\mathcal{C}$ are sequentially closed and $\mathcal{R}_+$-stable.
\end{asp}
It follows from \cite{CH11} that $\mathcal{A}:=\{\sigma'\pi:\pi\in\Lambda\}$ is sequentially closed and $\mathcal{R}_+$-\emph{stable} subset of $\mathcal{R}^n_+$. Writing $p_t=\sigma'_t\pi_t$, we also call such $p$ the investment strategy.

For an predictable process $u$ in $\mathbb{R}^{n}_+$, the distance between $u$ and $\mathcal{A}$ is a predictable process defined as$$\operatorname{dist}\left(\mathcal{A},u\right):=\operatorname{essinf}_{p\in\mathcal{A}}|u-p|.$$
The set $\Pi\left(\mathcal{A},u\right)$ consists of those elements in $\mathcal{A}$ at which the greatest lower bound with respect to the $\lambda\otimes\mathbb{P}\raisebox{0mm}{-}a.s.$ order is obtained
$$\Pi\left(\mathcal{A},u\right):=\left\{p\in\mathcal{A}:|u-p|=\operatorname{dist}\left(\mathcal{A},u\right)\right\}.$$
It is shown in Corollary 4.5 of \cite{CKV15} that $\Pi\left(\mathcal{A},u\right)$ is nonempty.

We enforce the following conditions on market coefficients.
\begin{asp}\label{para}
Each of $b,a,r,\mu,\sigma,\rho,\rho^\perp$ is continuous in $t$ and local  Lipschitz-continuous in domain $E$. $A:=aa'$ and $\Sigma:=\sigma\sigma'$ are both positive definite. $r\geq r_{\operatorname{min}}$ and $0<\mu'\Sigma^{-1}\mu\leq C_0$,  where $r_{\operatorname{min}}$ and $C_0$ are two constants. Without loss of generality, suppose $r_{min}<0$.
\end{asp}

Note that we work in a financial environment with a bounded from below interest rate $r$ and a bounded square of market risk price $\mu'\Sigma^{-1}\mu$, where both the risk premium $\mu$ and volatility $\sigma$ can be unbounded. Since $r$ is bounded from below, so is $r+\frac{1}{2\gamma}\mu'\Sigma^{-1}\mu$.

An agent must choose a consumption process $c$ and an investment strategy $p$ to invest in this financial market. Given an initial wealth $\omega$, the corresponding wealth process $\mathcal{W}^{c,p}$ is given by
$$d\mathcal{W}_t^{c,p}=\mathcal{W}_t^{c,p}\left(\left(r_t+p_t'\sigma_t'\Sigma_t^{-1}\mu_t\right)dt+p_t'dW_t^\rho\right)-c_tdt,\quad\mathcal{W}_0^{c,p}=\omega,$$
where $r_t,\sigma_t,\mu_t,\rho_t$ represents $r(t,X_t),\sigma(t,X_t),\mu(t,X_t),\rho(t,X_t)$ respectively.
A strategy $(c,p)$ is called \emph{admissible} if it belongs to
\begin{align}\label{gas}
\mathcal{S}_a=\left\{(c,p):c\in\mathcal{C}, p\in\mathcal{A}\ {\rm and} \ \left(\mathcal{W}^{c,p}\right)^{1-\gamma}e^Y {\rm\ is\ of\ class\ D}\right\},
\end{align}
where $Y$ is one solution to BSDE (\ref{mar}).

The agent wants to solve the maximization problem
\begin{align}\label{max}
V_0:=\sup_{(c,p)\in\mathcal{S}_a}V_0^{c,p}=\sup_{(c,p)\in\mathcal{S}_a}\mathbf{E}\left[\int_0^T f(c_s, V_s^c)ds+U(\mathcal{W}^{c,p}_T)\right].
\end{align}

\subsection{The martingale optimal principle}
For convenience, we do not put closed constraints on consumption process $c$ in Subsection 3.2 and 3.3. See Remark \ref{all} for corresponding results about $c$. In this case, the set of admissible strategies $\mathcal{S}_a$ has the following form
$$\mathcal{S}_a=\left\{(c,p):c\in\mathcal{C}_a, p\in\mathcal{A}\ {\rm and} \ \left(\mathcal{W}^{c,p}\right)^{1-\gamma}e^Y {\rm\ is\ of\ class\ D}\right\}.$$
We make the following assumption for investment strategies:
\begin{align}\label{bap}
{\rm there \ exists\  a\  bounded\  process\ }\underline{p}\ {\rm in}\ \mathcal{A}\ {\rm such\ that}\ |\underline{p}|\leq C_p\ {\rm where}\ C_p\ {\rm is\ a\ constant}.
\end{align}

In the following we suppress the supscript $(c,p)$ of $\mathcal{W}^{c,p}$. By the martingale optimal principle, we construct a so-called utility process
\begin{align}\label{upo}
G_t^{c,p}:=\frac{\mathcal{W}_t^{1-\gamma}}{1-\gamma}e^{Y_t}+\int_0^tf\left(c_s,\frac{\mathcal{W}_s^{1-\gamma}}{1-\gamma}e^{Y_s}\right)ds,\quad t\in[0,T],
\end{align}
where
\begin{align}\label{mar}
 Y_t=\int_t^TH(s,Y_s,Z_s)ds-\int_t^TZ_sdW_s,\quad t\in[0,T],
\end{align}
with
\begin{align}\label{generator-complete}
H(t,y,z)=&-\frac{\gamma(1-\gamma)}{2}\operatorname{dist}^2\left(\mathcal{A}_t, \frac{1}{\gamma}\sigma_t'\Sigma_t^{-1}(\mu_t+\sigma_t\rho_tz')\right)+z\left(\frac{1}{2}I_k+\frac{1-\gamma}{2\gamma}\rho_t'\sigma_t'\Sigma_t^{-1}\sigma_t\rho_t\right)z'\\\notag
&+\frac{1-\gamma}{\gamma}\mu_t'\Sigma_t^{-1}\sigma_t\rho_tz'+\frac{\theta}{\psi}\delta^\psi e^{-\frac{\psi}{\theta}y}+\frac{1-\gamma}{2\gamma}\mu_t'\Sigma_t^{-1}\mu_t+(1-\gamma) r_{t}-\delta \theta.
\end{align}
In fact, $H$ is chosen such that $G$ is a local martingale for candidate optimal strategy.   Comparing with \cite{X17},  BSDE \eqref{mar} derived by the martingale optimal principle is more complicated, which is involved a projection term on a closed set.  Fortunately, this term does not essentially change the quadratic structure of the BSDE.  Similar to \cite{X17}, the upper boundedness for the solution still holds under the parameter specification $\gamma>1$ and $\psi>1$.   For the convenience of the proof, we add the following Assumption \ref{IC}, making us to deal with the linear term of $z$ by Girsanov transformation and to derive the estimations of $Y$.

\begin{asp}\label{IC}
$$\frac{d\hat{\mathbb{P}}}{d\mathbb{P}}=\mathscr{E}\left(\int\frac{1-\gamma}{\gamma}\mu'\Sigma^{-1}\sigma\rho dW_s\right)_T$$ defines a new probability measure $\hat{\mathbb{P}}$ equivalent to $\mathbb{P}$. In addition (throughout the paper, $\hat{\mathbf{E}}(\cdot)$ denotes the expectation with respect to $\hat{\mathbb{P}}$), $\hat{\mathbf{E}}\left[\int_0^Tr_sds\right]<+\infty.$
\end{asp}

The following proposition shows that $Y$ is bounded from above, which is very important for our subsequent proofs.
\begin{pr}\label{Yestimate}
Let $\gamma, \psi>1$ and Assumptions \ref{ccc}, \ref{para} and \ref{IC} hold. Then BSDE (\ref{mar}) has at least one solution $(Y,Z)$ such that
\begin{align}
(1-\gamma)\hat{\mathbf{E}}_t\left[\int_t^Tr_sds\right]+C_2(t)\leq Y_t\leq C_1T, \quad \forall  t\in[0,T],
\end{align}
and $\hat{\mathbf{E}}\left(\int_0^T|Z_s|^2ds\right)<+\infty$, where
\begin{align*}
&C_1:=(1-\gamma)r_{min}-\delta\theta+2(C_0+\gamma(\gamma-1)C_p),\\
&C_2(t):=\left[\frac{\theta}{\psi}\delta^\psi\operatorname{exp}\left(-\frac{\psi}{\theta}C_1T\right)+\frac{1-\gamma}{2\gamma}C_0-\delta\theta\right](T-t).
\end{align*}
\end{pr}


Now we are ready to declare the main theorem in this subsection. Before that write $\mathcal{W}^{c^*,p^*}$ briefly as $\mathcal{W}^*$, where $c^*$ and $p^*$ are given by (\ref{os}).

\begin{thm}(Martingale optimal principle)\label{g}
Let Assumptions \ref{ccc}, \ref{para} and \ref{IC} hold. Suppose $(Y,Z)$ is a solution to BSDE (\ref{mar}). Then
\begin{enumerate}
\item[(i)] For any $(c,p)\in\mathcal{S}_a$, $G^{c,p}$ is a local supermartingale and
$$V_t^c\leq\frac{\mathcal{W}_t^{1-\gamma}}{1-\gamma}e^{Y_t},  \quad \forall t\in[0,T].$$
\item[(ii)] Denote
\begin{align}\label{os}
c^*=\hat{c}^*\mathcal{W}^*=\delta^\psi e^{-\frac{\psi}{\theta}Y}\mathcal{W}^*,\quad p^*\in\Pi\left(\mathcal{A}, \frac{1}{\gamma}\sigma'\Sigma^{-1}(\mu+\sigma\rho Z')\right),
\end{align}
where $\mathcal{W}^{*}$ is the corresponding optimal wealth process. If $(c^*,p^*)\in\mathcal{S}_a$, then $G^{c^*,p^*}$ is a local martingale and $(c^*,p^*)$ is the optimal strategy for problem (\ref{max}). Moreover, for the initial wealth $\omega$, the optimal utility is given by
$$V_0^{c^*}=\frac{\omega^{1-\gamma}}{1-\gamma}e^{Y_0}.$$
\end{enumerate}

\end{thm}

\begin{rem}
The martingale optimal principle here is different from the one in \cite{HIM05}. Thanks to the specific structure of Epstein-Zin utility, it is sufficient to obtain the optimality of utility and strategy that $G^{c,p}$ is a local supermartingale and $G^{c^*,p^*}$ is a local martingale.
\end{rem}

\subsection{The verification theorem}
In order to verify the martingale optimal principle, it is necessary to verify the stochastic exponential
\begin{align}\label{mse}
M:=\mathscr{E}\left(\int(1-\gamma)(p^*)'dW^\rho+\int ZdW\right)
\end{align}
is a martingale under $\mathbb{P}$. We introduce an operator $\mathfrak{F}$ of a Lyapunov function $\phi$ to realize this goal. We borrowed this method from Chapter 10 of \cite{SV06} and \cite{X17}.  The readers can also refer to \cite{RX17}  for more applications.
\begin{asp}\label{Lfo}
Suppose that $\frac{1}{2}+\frac{2(1-\gamma)}{\gamma}\left|\sigma'\Sigma^{-1}\sigma\rho\right|\left|\rho\right|>0$ and $\phi\in C^2(E)$ is a Lyapunov function which satisfies following properties
\begin{enumerate}
\item[(i)] $\lim_{n\to\infty}\inf_{x\in E\backslash E_n}\phi(x)=\infty$, where $(E_n)$ is a sequence of open domains in $E$ satisfying $\cup_nE_n=E$, $\overline{E}_n$ compact and $\overline{E}_n\subseteq E_{n+1}$ for every $n$.
\item[(ii)] The operator
\begin{align}\label{Lyapunov}
\mathfrak{F}\left(\phi\right):=&b'\nabla\phi+\frac{1}{2}\sum_{i,j=1}^{k}A_{ij}\partial^2_{x_ix_j}\phi-\frac{2(1-\gamma)}{\gamma}\left|\sigma'\Sigma^{-1}\mu\right|\left|\rho a'\nabla\phi\right|-(1-\gamma)C_p\left|\rho a'\nabla\phi\right|\notag\\
&+\frac{1}{4}\frac{\left((1-\gamma)C_p|\rho|+\frac{2(1-\gamma)}{\gamma}\left|\sigma'\Sigma^{-1}\mu\right|\left|\rho\right|-\left|\nabla\phi'a\right|+\frac{2(1-\gamma)}{\gamma}\left|\sigma'\Sigma^{-1}\sigma\rho\right|\left|\rho a'\nabla\phi\right|\right)^2}{\frac{1}{2}+\frac{2(1-\gamma)}{\gamma}\left|\sigma'\Sigma^{-1}\sigma\rho\right|\left|\rho\right|}
\end{align}
is bounded from above on $E$.
\end{enumerate}
\end{asp}
\begin{rem}Our operator $\mathfrak{F}\left(\cdot\right)$ is much different from \cite{X17}.
Since we impose closed constraints on the set of admissible strategies, which causes the candidate optimal investment strategy \eqref{os} cannot be expressed by $Z$ explicitly.  After subtle estimations and careful calculations, we delicately choose a suitable Lyapunov function to overcome the difficulties. In Section 4, we will give some examples and specific Lyapunov functions $\phi$ satisfying Assumption \ref{Lfo}. Besides, the parameter condition $\frac{1}{2}+\frac{2(1-\gamma)}{\gamma}\left|\sigma'\Sigma^{-1}\sigma\rho\right|\left|\rho\right|>0$ ensures (\ref{Lyapunov}) is well-defined, and it always holds when $|\rho|$ is small enough.
\end{rem}

Before giving the verification theorem, the following assumption ensures the integrability under a risk-neutral measure.
\begin{asp}\label{rnm}
There exists a risk-neutral measure $\mathbb{P}_0$ denoted by $$\frac{d\mathbb{P}_0}{d\mathbb{P}}=\mathscr{E}\left(\int-\lambda dW_s^\rho\right)_T,$$
where $\lambda:\mathbb{R}_+\times E\to\mathbb{R}^{1\times n}$ is defined by $\mu'=\lambda\sigma'$. Furthermore,
$$\mathbf{E}^0\left[\operatorname{exp}\left((\psi-1)\int_0^Tr_s^+ds\right)\mathscr{E}\left(\int\lambda_s dW_s^0\right)^\psi_T\right]<+\infty,$$
where $\mathbf{E}^0(\cdot)$ denotes the expectation under $\mathbb{P}_0$ and $W^0$ is a Brownian motion under $\mathbb{P}_0$.
\end{asp}

\begin{rem}\label{degen}
If $r$ and $\lambda$ are both bounded, the above assumption holds automatically. If $\sigma$ is symmetric,  the risk-neutral measure $\mathbb{P}_0$ exists due to the boundedness of $\mu'\Sigma^{-1}\mu$ from Assumption \ref{para}. The integrability condition in Assumption \ref{rnm} can degenerate into a simpler one as follows: for sufficiently small $\epsilon>0$,
$$\mathbf{E}^0\left[\operatorname{exp}\left((1+\epsilon)(\psi-1)\int_0^Tr_s^+ds\right)\right]<+\infty.$$
\end{rem}

Then we can give our main result in this paper.

\begin{thm}\label{main}
Let $\gamma$, $\psi>1$. Suppose Assumptions \ref{ccc}, \ref{para}, \ref{IC}, \ref{Lfo} and \ref{rnm} hold. Then $c^*$ and $p^*$ defined in (\ref{os}) maximize the Epstein-Zin utility among all admissible strategies.
\end{thm}

\begin{cor}\label{last}
Suppose $\gamma$, $\psi>1$ and all market parameters are bounded. Then $c^*$ and $p^*$ defined in (\ref{os}) maximize the Epstein-Zin utility among all admissible strategies. Moreover, for any initial wealth $\omega$, the optimal Epstein-Zin utility is given by $$V_0^{c^*}=\frac{\omega^{1-\gamma}}{1-\gamma}e^{Y_0},$$
where $Y$ is one solution to BSDE (\ref{mar}).
\end{cor}

\noindent {\bf Proof.}  When all market parameters are bounded, Assumptions \ref{para}, \ref{IC} and \ref{rnm} hold obviously. Besides, Assumption \ref{Lfo} is not needed. Indeed, when all parameters are bounded, $Y$ is bounded, and the stochastic exponential $M$ can be proved to be a martingale by using the BMO argument directly.
\qed

\begin{rem}
Corollary \ref{last} provides an optimization result under bounded market parameters with Epstein-Zin type of recursive utilities. When $\gamma=\frac{1}{\psi}$, $V_0^c$ degenerates into time-separable utilities.  One can refer to \cite{HIM05} and \cite{CH11} for more details.
\end{rem}

\begin{rem}\label{all}
Now we present results for the consumption process $c$ with closed constraints. We make a similar assumption on $\mathcal{C}$: there exists a bounded process $\underline{c}\in\mathcal{C}$. (When there is no constraints on $\mathcal{C}$, this condition is automatically satisfied since $0\in\mathcal{C}$.)

Recall that the admissible strategy is in (\ref{gas}) where
\begin{align*}
 Y_t=\int_t^TH(s,Y_s,Z_s)ds-\int_t^TZ_sdW_s,\quad t\in[0,T],
\end{align*}
with
\begin{align*}
&H(t,y,z)\\
=&-\frac{\gamma(1-\gamma)}{2}\operatorname{dist}^2\left(\mathcal{A}_t, \frac{1}{\gamma}\sigma_t'\Sigma_t^{-1}(\mu_t+\sigma_t\rho_tz')\right)+z\left(\frac{1}{2}I_k+\frac{1-\gamma}{2\gamma}\rho_t'\sigma_t'\Sigma_t^{-1}\sigma_t\rho_t\right)z'\\\notag
&+\frac{1-\gamma}{\gamma}\mu_t'\Sigma_t^{-1}\sigma_t\rho_tz'+\inf _{\hat{c}\in\mathcal{C}}\left(-(1-\gamma) \hat{c}+\delta \theta e^{-\frac{1}{\theta} y} \hat{c}^{1-\frac{1}{\psi}}\right)+\frac{1-\gamma}{2\gamma}\mu_t'\Sigma_t^{-1}\mu_t+(1-\gamma) r_{t}-\delta \theta.
\end{align*}
Due to Assumption \ref{ccc}, we have the following estimation
\begin{align}\label{ce}
\frac{\theta}{\psi}\delta^\psi e^{-\frac{\psi}{\theta}Y}&=\inf _{\hat{c}\in \mathcal{R}_+}\left(-(1-\gamma) \hat{c}+\delta \theta e^{-\frac{1}{\theta} Y} \hat{c}^{1-\frac{1}{\psi}}\right)\notag\\
&\leq\inf _{\hat{c}\in\mathcal{C}}\left(-(1-\gamma) \hat{c}+\delta \theta e^{-\frac{1}{\theta} Y} \hat{c}^{1-\frac{1}{\psi}}\right)\notag\\
&\leq-(1-\gamma) \underline{c}+\delta \theta e^{-\frac{1}{\theta} Y} \underline{c}^{1-\frac{1}{\psi}}.
\end{align}

Using (\ref{ce}), we can also construct a supersolution $\overline{Y}$ and a subsolution $\underline{Y}$ similar to Proposition \ref{Yestimate}. Then the localization technique in \cite{BH06} derives that $Y$ is bounded from above and $Z$ is square integrable.

When Assumptions \ref{ccc}, \ref{para}, \ref{IC}, \ref{Lfo} and \ref{rnm} hold, we can obtain Theorem \ref{main} as well, where the optimal strategy can be denoted by
\begin{align}\label{gs2}
\hat{c}^*\in\operatorname{arg\ inf}_{\hat{c}\in\mathcal{C}}\left(-(1-\gamma) \hat{c}+\delta \theta e^{-\frac{1}{\theta}Y} \hat{c}^{1-\frac{1}{\psi}}\right),\quad p^*\in\Pi\left(\mathcal{A}, \frac{1}{\gamma}\sigma'\Sigma^{-1}(\mu+\sigma\rho Z')\right).
\end{align}

In fact, it suffices to prove the admissibility of $\hat{c}^*$. Noting that inequality \eqref{ce} implies that
$\inf _{\hat{c}\in\mathcal{C}}\left(-(1-\gamma) \hat{c}+\delta \theta e^{-\frac{1}{\theta} Y} \hat{c}^{1-\frac{1}{\psi}}\right)$ is bounded. We can also get the boundedness of  $\hat{c}^{*}$ in $\operatorname{arg\ inf}_{\hat{v}\in\mathcal{C}}\left(-(1-\gamma) \hat{v}+\delta \theta e^{-\frac{1}{\theta}Y} \hat{v}^{1-\frac{1}{\psi}}\right)$.  Indeed, 
by \eqref{ce},  it derives 
$$\frac{\theta}{\psi}\delta^\psi e^{-\frac{\psi}{\theta}Y}\leq (\gamma-1)\hat{c}^*+\delta\theta e^{-\frac{Y}{\theta}}(\hat{c}^*)^{1-\frac{1}{\psi}}\leq (\gamma-1)\underline{c}.$$Due to the fact that $\theta<0$, $\gamma>1$ and $\psi>1$ and the upper boundedness of $Y$,  it implies $(\gamma-1)\hat{c}^*+\delta\theta e^{-\frac{Y}{\theta}}(\hat{c}^*)^{1-\frac{1}{\psi}}=(\hat{c}^*)^{1-\frac{1}{\psi}}[(\gamma-1)(\hat{c}^*)^{1/\psi}+\delta\theta e^{-\frac{Y}{\theta}}]$ is a bounded process.  Hence $\hat{c}^*$ is bounded.

Similar to the proof of Theorem \ref{main}, we can show
$$\mathbf{E}\left[\int_0^{T}e^{-\delta s}\left(c^*_s\right)^{1-\frac{1}{\psi}}ds\right]<+\infty,\quad \mathbf{E}\left[\left(c^*_T\right)^{1-\gamma}\right]<+\infty,$$
and
\begin{align*}
\left(\mathcal{W}_t^*\right)^{1-\gamma}e^{Y_t}=\omega^{1-\gamma}e^{Y_0}\operatorname{exp}\left(-\int_0^t\left(\delta\theta(\hat{c}^*_s)^{1-\frac{1}{\psi}}e^{-\frac{1}{\theta}Y_s}-\delta\theta\right)ds\right)M_t,
\end{align*}
where $c^*=\hat{c}^*\mathcal{W}^*$, $M$ is of class D and other terms are bounded. Hence $\left(\mathcal{W}^*\right)^{1-\gamma}e^Y$ is of class D on [0,T]. Then $\hat{c}^*$ is admissible. Therefore, we obtain the fact that (\ref{gs2}) is the optimal strategy.
\end{rem}
\begin{rem} 
In particular,  we can also impose $0\in\mathcal{A}$ and $0\in\mathcal{C}$ to replace with
 the existence of two bounded processes $\underline{p}\in \mathcal{A}$ and  $\underline{c}\in \mathcal{C}$.  
When $0\in \mathcal{A}$, although there may exist a consumption strategy $c$ such that $(0, c)\in\mathcal{S}_{a}$,  we just add conditions on $\mathcal{A}$ or $\mathcal{C}$,  not asking for a pair of strategy $(\underline{p},c)$ or $(p,\underline{c})$ in $\mathcal{S}_{a}$.  The nonempty property of $\mathcal{S}_{a}$ is guaranteed by the existence of optimal strategy $(c^{*},p^{*})$. 
\end{rem}

\section{Numerical examples}

This section provides three specific numerical examples to illustrate the influence of the constraints and comparative statics analysis on optimal strategies. The Black-Scholes model, linear diffusion model and Heston stochastic volatility model  are examined respectively. We adopt Markovian quantization method to simulate the state process $X$ and  use  Monte-Carlo approximation method proposed by \cite{CR16} to simulate BSDE (\ref{mar}).

For convenience, we only take one dimensional case into account, and only consider closed constraints on the investment strategy $\pi$.  

\subsection{Black-Scholes model}
This subsection considers a much simple example, where the risky asset follows the classical Black-Scholes model and each parameter is constant.  In particular, specify the following parameters
\begin{align}\label{parameters0}
r=0.03,  \mu=0.05,  \sigma=0.17.
\end{align}
The assignment of parameters can be found in \cite{BMZ14}. We intend to compare the optimal portfolio $\pi^*$ and the optimal consumption wealth ratio $\hat{c}^*$ for CRRA utility with constraints (\cite{CH11}), and Epstein-Zin utility without constraints (\cite{X17}) and Epstein-Zin utility with constraints (this paper) by numerical simulations respectively.

\begin{figure}[htbp]
\subfigure[]{\includegraphics[width=7.2cm,height=3.6cm]{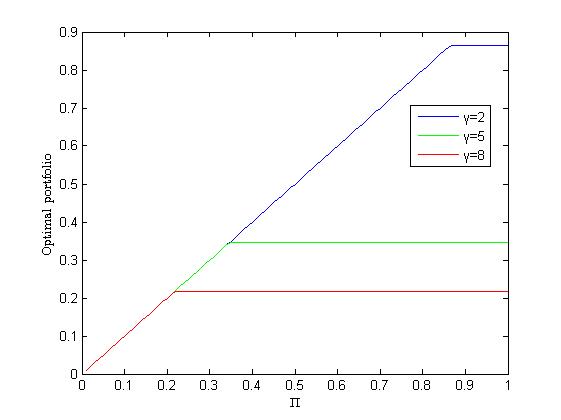}
}
\hspace{2mm}
\subfigure[]{\includegraphics[width=7.2cm,height=3.6cm]{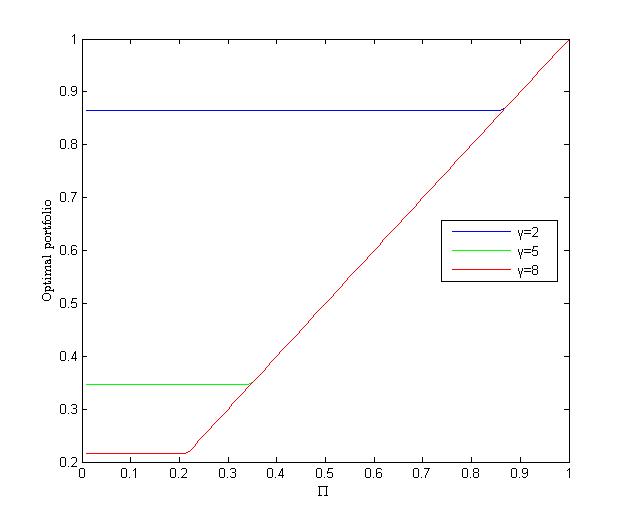}
}
\caption{All Parameter values are given in (\ref{parameters0}). $\gamma=2$ and $\psi=1.2$ without special instructions. The discount rate $\delta=0.08$. They are both time-0 values for a problem with time horizon $T=30$ years.  }
\end{figure}

Figure 1 shows the impact of the constraint on the optimal portfolio $\pi^*$ when we let $\Pi$ change with constraints $\pi\in[0,\Pi]$ and $\pi\in[\Pi,1]$ respectively. In Figure 1(a), the optimal portfolio $\pi^*$ increases when $\Pi$ increases until the optimal portfolio on $[0,1]$ is reached. After this $\pi^*$ does not change. In Figure 1(b), $\pi^*$ keeps the optimal portfolio on $[0,1]$ before this value and then increases as $\Pi$ increases. We can also get that the optimal portfolio $\pi^*$ on $[0,1]$ increases when $\gamma$ increases in both subfigures.

\begin{figure}[htbp]
\subfigure[]{\includegraphics[width=6.8cm,height=3.6cm]{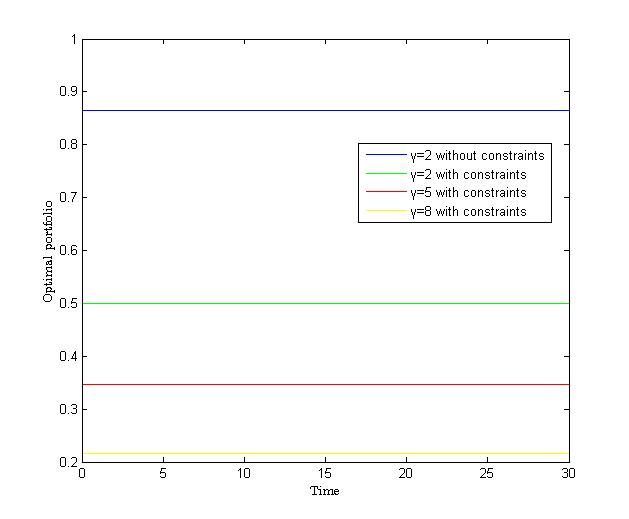}
}
\hspace{2mm}
\subfigure[]{\includegraphics[width=7.8cm,height=3.6cm]{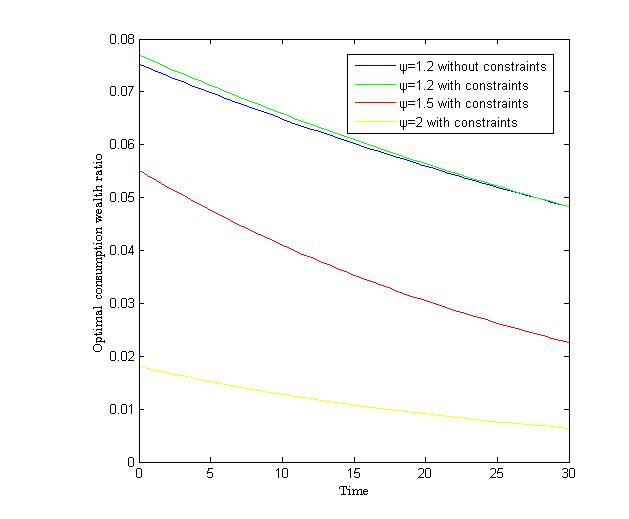}
}
\caption{All Parameter values are given in (\ref{parameters0}). $\gamma=2$ and $\psi=1.2$ without special instructions. The time horizon is $T=30$ years.  The discount rate $\delta=0.08$.}
\end{figure}

Then we consider the closed constraint $\pi\in[0,0.5]$.  Figure 2 indicates that  the closed constraint makes the optimal portfolio $\pi^*$ smaller and the optimal consumption wealth ratio $\hat{c}^*$ larger. But the change of $\hat{c}^*$ is not obvious. Figure 2(a) shows that $\pi^*$ is constant with respect to $t$.  Besides,  $\pi^*$ decreases as $\gamma$ increases and $\hat{c}^*$ decreases as $\psi$ increases when we consider constraints on $\pi^*$.

\subsection{Linear diffusion model}
The linear diffusion model, whose interest rate and the excess return of risky assets are linear functions. This model has been investigated in \cite{CV99} for recursive utilities. In our situation,  we truncate these linear functions to satisfy our assumptions as follows
\[
\begin{dcases}
dX_t=-bX_tdt+adW_t,\\
dS_t=S_t\left[\left(r(X_t)+\mu(X_t)\right)dt+\sigma dW^\rho_t\right],
\end{dcases}
\]
where $r(x)=r_0+r_1((-100)\vee x)$, $\mu(x)=\sigma(\lambda_0+\lambda_1((-100)\vee x\wedge100))$, given $b,a,\sigma,r_0,r_1,\lambda_0,\lambda_1,\rho,\rho^\perp\in\mathbb{R}$. These parameters satisfy the conditions in the following proposition.

\begin{pr}\label{exp2}
Suppose $\gamma,\psi>1$ and
\begin{enumerate}
\item[(i)] $b>0$, $a>0$,
\item[(ii)] $r_1>0$,
\item[(iii)] $\frac{a^2}{b}<\frac{1+\frac{4(1-\gamma)}{\gamma}\rho^2}{\left(\frac{2(1-\gamma)}{\gamma}\rho^2-1\right)^2}$,
\item[(iv)] $(\psi-1)r_1<\frac{(b-(\psi-1)a\lambda_1\rho)^2}{2a^2}$.
\end{enumerate}
Then Theorem \ref{main} holds.
\end{pr}

This assumption ensures that X takes values in $E=\mathbb{R}$. The  values of parameters are
\begin{align}\label{parameters2}
r=0.0014,\lambda_0=0.05,\lambda_1=1,\sigma=0.0436,
b=0.0226,a=0.0189,\rho=-0.935,\delta=0.0052,
\end{align}
which can be found in \cite{W02}. We consider the closed constraint $\pi\in[0,0.5]$.

\begin{figure}[htbp]
\subfigure[]{\includegraphics[width=6.8cm,height=3.6cm]{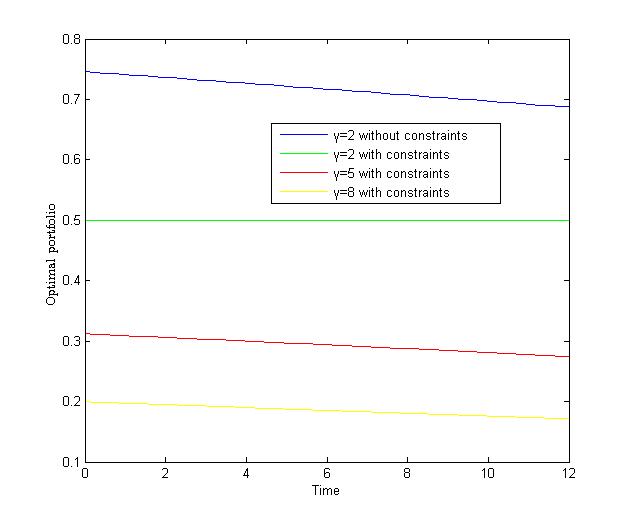}
}
\hspace{2mm}
\subfigure[]{\includegraphics[width=7.8cm,height=3.6cm]{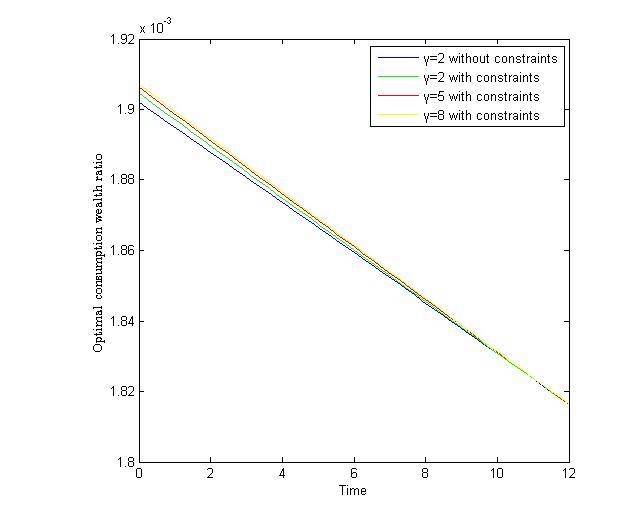}
}
\caption{Each figure uses the parameters given in (\ref{parameters2}), $T=12$ month. Without special instructions, $\gamma=2$ and $\psi=1.2$. }
\end{figure}

Figure 3 compares the optimal portfolio $\pi^*$ and the optimal consumption wealth ratio $\hat{c}^*$ for portfolio strategies with and without closed constraints. Our numerical results show that the closed constraints has a significant impact on $\pi^{*}$ but hardly change $\hat{c}^*$. Two panels also illustrate the effects on $\pi^*$ and $\hat{c}^*$ for different risk aversion coefficients $\gamma$. When the risk aversion increases, the optimal portfolio is decreasing while the consumption wealth ratio is increasing.

\begin{figure}[htbp]
\subfigure[]{\includegraphics[width=7.2cm,height=3.6cm]{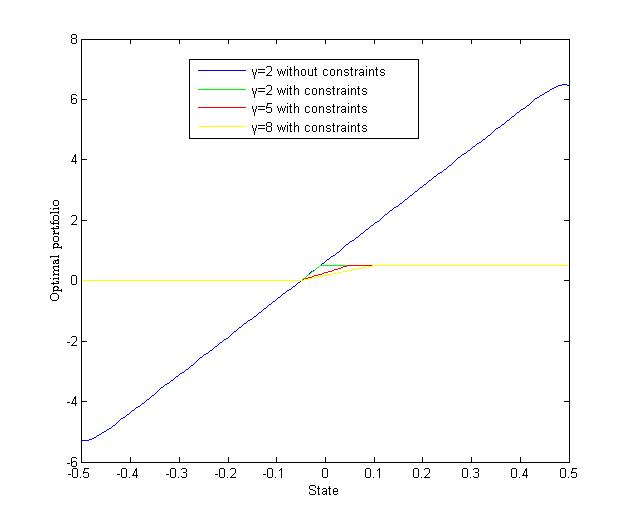}
}
\hspace{2mm}
\subfigure[]{\includegraphics[width=7.2cm,height=3.6cm]{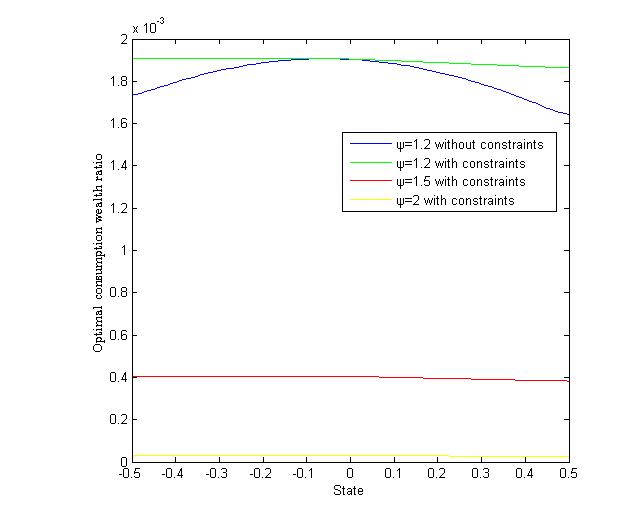}
}
\caption{All Parameter values are given in (\ref{parameters0}). $\gamma=2$ and $\psi=1.2$ without special instructions. They are both time-0 values for a problem with the discount rate $\delta=0.08$. }
\end{figure}

Figure 4 compares the optimal portfolio $\pi^*$ and the optimal consumption wealth ratio $\hat{c}^*$ for portfolio strategies with and without closed constraints with respect to the state variable $X$. When we impose constraint $\pi\in[0,0.5]$, the part of $\pi^*$ greater than 0.5 takes 0.5 and the part less than 0 takes 0. In this case $\pi^*$ grows slower when $\gamma$ increases. It is shown in Figure 4(b) that $\hat{c}^*$ get smaller with respect to $\psi$ and is affected by state variable $X$ not obviously.

\subsection{Stochastic volatility model}
The state process $X$ is following a square root process, which is suggested by Heston and further investigated by \cite{CV05},
\[
\begin{dcases}
dX_t=b(l-X_t)dt+a\sqrt{X_t}dW_t,\\
dS_t=S_t\left[\left(r(X_t)+\mu(X_t)\right)dt+\sigma (X_t)dW^\rho_t\right],
\end{dcases}
\]
where $r(x)=r_0+r_1x$, $\mu(x)=\sigma\lambda x$, $\sigma(x)=\sigma x$, given $b,l,a,r_0,r_1,\sigma,\lambda,\rho,\rho^\perp\in\mathbb{R}$. These coefficients are subject to some restrictions in the following proposition.

\begin{pr}\label{exp1}
Suppose $\gamma,\psi>1$ and
\begin{enumerate}[{\rm(i)}]
\item $b,l,a,r_1,\sigma,\lambda>0, bl>\frac{1}{2}a^2$,
\item $\frac{1}{2}+\frac{2(1-\gamma)}{\gamma}|\rho|^2>0$,
\item $(\psi-1)r_1<\frac{b^2}{2a^2}$.
\end{enumerate}
Then Theorem \ref{main} holds.
\end{pr}

Let us stress here that $X$ takes values in $(0,+\infty)$ with $E=(0,+\infty)$ and $r$ is bounded from below by $r_0$. Conditions in Proposition \ref{exp1} already contain many empirically specifications in \cite{LP03}, where the values of these market parameters are
\begin{align}\label{parameters1}
r=0.05,\ \lambda=0.47,\ l=0.0225,\ \sigma=1,\ b=5,\ a=0.25,\ \rho=-0.5,\ \delta=0.08.
\end{align}
Since the optimal portfolio $\pi^*$ without constraints lies in $[0,0.5]$, we choose portfolio constraint $\pi\in[0,0.1]$ in this model.

\begin{figure}[htbp]
\subfigure[]{\includegraphics[width=6.8cm,height=3.6cm]{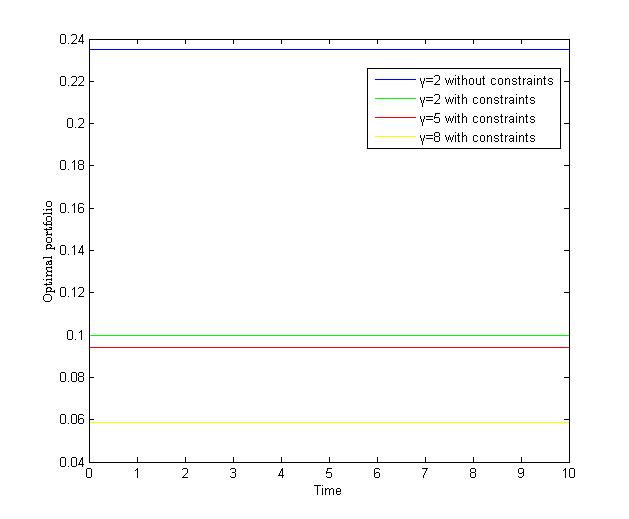}
}
\hspace{2mm}
\subfigure[]{\includegraphics[width=7.8cm,height=3.6cm]{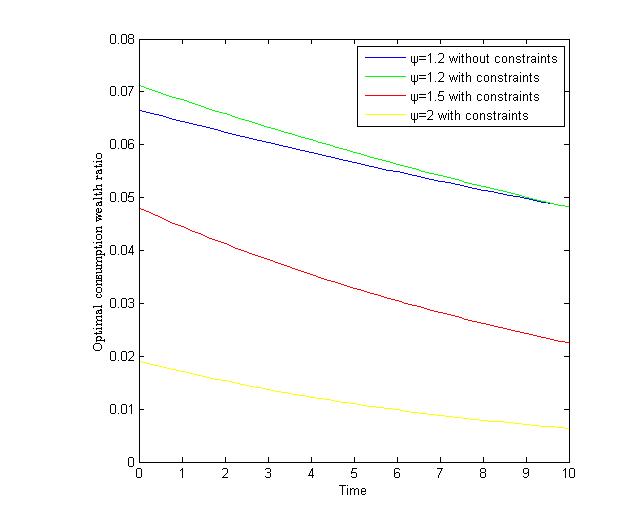}
}
\caption{Parameter values are given in (\ref{parameters1}). Time horizon $T=10$ years. Without special instructions, $\gamma=2$ and $\psi=1.2$. }
\end{figure}

\linespread{0}

Figure 5 compares the optimal investment fraction $\pi^*$ and the optimal consumption wealth ratio $\hat{c}^*$ with respect to time $t$ for different values of the risk aversion $\gamma$ and the EIS $\psi$. It can be seen that $\gamma$ changes $\pi^*$ observably. Intuitively, an agent with larger risk aversion is more conservative and he will invest fewer wealth in risky asset.  In Figure 5(b),
the consumption wealth ratio is decreasing with respect to $\psi$.  The figure also displays $\pi^*$ and  $\hat{c}^*$ for portfolio strategies with and without closed constraints when $\gamma=2$ and $\psi=1.2$. Once we constrain the portfolio strategy, $\pi^*$ get smaller and  $\hat{c}^*$ get bigger.

\begin{figure}[htbp]
\subfigure[]{\includegraphics[width=6.8cm,height=3.6cm]{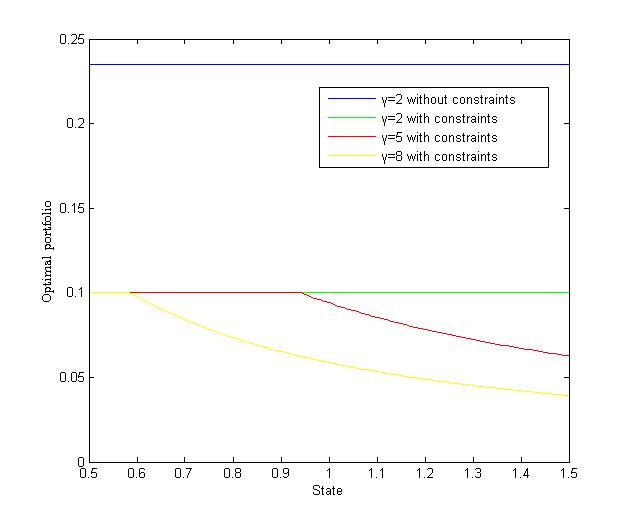}
}
\hspace{2mm}
\subfigure[]{\includegraphics[width=7.8cm,height=3.6cm]{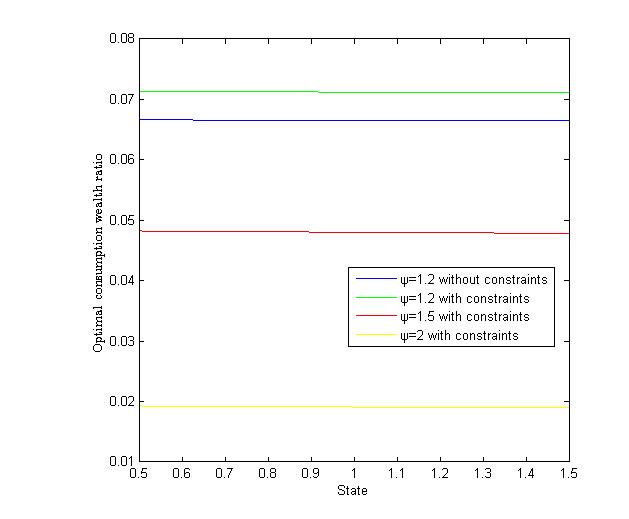}
}
\caption{Parameter values are given in (\ref{parameters1}). Time horizon $T=10$ years. Without special instructions, $\gamma=2$ and $\psi=1.2$. }
\end{figure}

Figure 6 shows the optimal portfolio $\pi^*$ and the optimal consumption wealth ratio $\hat{c}^*$ with respect to the state variable $X$. Figure 6(b) depicts that $\hat{c}^*$ hardly changes depending on $X$. It becomes more complicated for $\pi^*$ as shown in Figure 6(a). When $\gamma=2$, $\pi^*$ remains 0.1; when $\gamma=5$ and 8, $\pi^*$ firstly stays constant and then decreases slowly.

\section{The Proofs}

\subsection{Proof of Proposition \ref{Yestimate}}
By Assumption \ref{IC}, it implies that $\hat{W}:=W-\int_0^\cdot\frac{1-\gamma}{\gamma}\rho_s'\sigma_s'\Sigma_s^{-1}\mu_s ds$ is a Brownian motion under the new probability measure $\hat{\mathbb{P}}$. BSDE (\ref{mar}) then can be rewritten under $\hat{\mathbb{P}}$ as follows
\begin{align}\label{mar-change}
\mathcal{Y}_t=\xi+\int_t^T\mathcal{H}(s,\mathcal{Y}_s,Z_s)ds-\int_t^TZ_sd\hat{W}_s,\quad t\in[0,T],
\end{align}
where $$\xi=\int_0^T\left(\frac{1-\gamma}{2\gamma}\mu_s'\Sigma_s^{-1}\mu_s+(1-\gamma) r_{s}-\delta \theta\right)ds,$$ $$\mathcal{Y}_t=Y_t+\int_0^t\left(\frac{1-\gamma}{2\gamma}\mu_s'\Sigma_s^{-1}\mu_s+(1-\gamma) r_{s}-\delta \theta\right)ds$$ and
\begin{align*}
\mathcal{H}(t,y,z)=&-\frac{\gamma(1-\gamma)}{2}\operatorname{dist}^2\left(\mathcal{A}_t, \frac{1}{\gamma}\sigma_t'\Sigma_t^{-1}(\mu_t+\sigma_t\rho_tz')\right)+z\big(\frac{1}{2}I_k+\frac{1-\gamma}{2\gamma}\rho_t'\sigma_t'\Sigma_t^{-1}\sigma_t\rho_t\big)z'\\
&+\frac{\theta}{\psi}\delta^\psi e^{-\frac{\psi}{\theta}y}\operatorname{exp}\left({\frac{\psi}{\theta}\int_0^t\left(\frac{1-\gamma}{2\gamma}\mu_s'\Sigma_s^{-1}\mu_s+(1-\gamma) r_{s}-\delta \theta\right)ds}\right).
\end{align*}

Similar to Proposition 2.9 in \cite{X17},  by the localization technique in \cite{BH06},  for $m\geq1$, we introduce $(\mathcal{Y}^m, Z^m)$ as the  truncated version of BSDE (\ref{mar-change}),
\begin{align}\label{mar-truncated}
\mathcal{Y}_t^m=\xi^m+\int_t^T\mathcal{H}^m(s,\mathcal{Y}_s^m,Z_s^m)ds-\int_t^TZ_s^md\hat{W}_s,\quad t\in[0,T],
\end{align}
where $\xi^m=\int_0^T\left(\frac{1-\gamma}{2\gamma}\mu_s'\Sigma_s^{-1}\mu_s+(1-\gamma)(r_{s}\wedge m)-\delta \theta\right)ds$ and
\begin{align*}
\mathcal{H}^m(t,y,z)=&-\frac{\gamma(1-\gamma)}{2}\operatorname{dist}^2\left(\mathcal{A}_t, \frac{1}{\gamma}\sigma_t'\Sigma_t^{-1}(\mu_t+\sigma_t\rho_tz')\right)+z\big(\frac{1}{2}I_k+\frac{1-\gamma}{2\gamma}\rho_t'\sigma_t'\Sigma_t^{-1}\sigma_t\rho_t\big)z'\\ \notag
&+\frac{\theta}{\psi}\delta^\psi \left(e^{-\frac{\psi}{\theta}y}\wedge m\right)\operatorname{exp}\left({\frac{\psi}{\theta}\int_0^t\left(\frac{1-\gamma}{2\gamma}\mu_s'\Sigma_s^{-1}\mu_s+(1-\gamma)(r_{s}\wedge m)-\delta \theta\right)ds}\right).
\end{align*}
Since the eigenvalues of $\sigma'\Sigma^{-1}\sigma$ are either 0 or 1, we obtain $0\leq\rho'\sigma'\Sigma^{-1}\sigma\rho\leq\rho'\rho\leq I_k$. Then we get the following two inequalities
\begin{align}\label{zi}
\frac{1}{2\gamma}I_k\leq\frac{1}{2}I_k+\frac{1-\gamma}{2\gamma}\rho'\sigma'\Sigma^{-1}\sigma\rho\leq\frac{1}{2}I_k
\end{align}
and
\begin{align}\label{quadratic}
\operatorname{dist}^2\left(\mathcal{A}_t, \frac{1}{\gamma}\sigma_t'\Sigma_t^{-1}(\mu_t+\sigma_t\rho_tz')\right)=&\operatorname{min}_{p\in \mathcal{A}}\left|p-\frac{1}{\gamma}\sigma_t'\Sigma_t^{-1}(\mu_t+\sigma_t\rho_tz')\right|^2\notag\\
\leq&\frac{4}{\gamma^2}\mu_t'\Sigma^{-1}_t\mu_t+\frac{4}{\gamma^2}z\rho_t'\sigma_t'\Sigma_t^{-1}\sigma_t\rho_tz'+4\operatorname{min}_{p\in\mathcal{A}}|p|^2\notag\\
\leq&\frac{4}{\gamma^2}C_0+4C_p+\frac{4}{\gamma^2}zz',
\end{align}
where the last inequality in \eqref{quadratic} is deduced from Assumptions \ref{ccc}-\ref{para}, \eqref{bap} and \eqref{zi}.

According to \eqref{zi}, \eqref{bap} and the fact that $\theta<0$, it implies that
\begin{align}\label{mq}
\mathcal{H}^m(t,y,z)&\leq\frac{5}{2}zz'+2C_0+2\gamma(\gamma-1)C_p.
\end{align}
Thus $\mathcal{H}^m$ is Lipschitz continuous in $y$ with quadratic growth in $z$. It follows from Theorem 2.3 of \cite{K00} that BSDE (\ref{mar-truncated}) admits a solution  $(\mathcal{Y}^m, Z^m)\in\mathcal{S}^\infty\times\mathcal{M}^2$, where ${S}^\infty$ denotes the set of 1-dimensional continuous adapted processes $Y$ such that $||\sup_{0\leq s\leq T}|Y_s|||_{\infty}<+\infty$,\ and $\mathcal{M}^2$ denotes the set of predictable processes $Z$ such that $\hat{\mathbf{E}}\left(\int_0^T|Z_s|^2ds\right)<+\infty$.

Next, we will find a priori bound for $\mathcal{Y}^m$ independent of $m$. By \eqref{mq}, we consider
\begin{align*}
\overline{Y}_t^m=\xi^m+\int_t^T\frac{5}{2}{\overline{Z}_s^m}{\overline{Z}_s^m}'+2C_{0}+2\gamma(\gamma-1)C_pds-\int_t^T{\overline{Z}_s^m}d\hat{W}_s,  \quad  t\in[0,T].
\end{align*}
The explicit solution $\overline{Y}_t^m=\frac{1}{5}\ln\hat{\mathbf{E}}_t\left[\operatorname{exp}\left(5\xi^m+10(C_0+\gamma(\gamma-1)C_p)T\right)\right]-2(C_0+\gamma(\gamma-1)C_p)t.$
Since $r\geq r_{\operatorname{min}}$ and $\gamma>1$, then
$\frac{1-\gamma}{2\gamma}\mu_s'\Sigma_s^{-1}\mu_s+(1-\gamma)(r_{s}\wedge m)\leq(1-\gamma)r_{\operatorname{min}}.$
Thus,
\begin{align*}
\overline{Y}_t^m&\leq[(1-\gamma)r_{\operatorname{min}}-\delta\theta+2(C_0+\gamma(\gamma-1)C_p)]T-2(C_0+\gamma(\gamma-1)C_p)t \leq C_1T
\end{align*}
and
\begin{align}\label{fanxiang}
&\overline{Y}_t^m-\int_0^t\left(\frac{1-\gamma}{2\gamma}\mu_s'\Sigma_s^{-1}\mu_s+(1-\gamma)(r_{s}\wedge m)-\delta \theta\right)ds\notag\\
=& \frac{1}{5}\ln\hat{\mathbf{E}}_t\left[\operatorname{exp}\left(5\int_t^T\left(\frac{1-\gamma}{2\gamma}\mu_s'\Sigma_s^{-1}\mu_s+(1-\gamma)(r_{s}\wedge m)-\delta \theta\right)ds+10(C_0+\gamma(\gamma-1)C_p)T\right)\right]\notag\\
&-2(C_0+\gamma(\gamma-1)C_p)t\notag\\
\leq& \frac{1}{5}\ln\hat{\mathbf{E}}_t\left[\operatorname{exp}\left(5[(1-\gamma)r_{\operatorname{min}}-\delta\theta](T-t)+10(C_0+\gamma(\gamma-1)C_p)T\right)\right]-2(C_0+\gamma(\gamma-1)C_p)t\notag\\
=&[(1-\gamma)r_{\operatorname{min}}-\delta\theta+2(C_0+\gamma(\gamma-1)C_p)](T-t)\nonumber\\
\leq & C_1T.
\end{align}
Since $\theta<0$ and $\gamma>1$, by  inequality (\ref{fanxiang}), it derives that
\begin{align*}
&\mathcal{H}^m(t,\overline{Y}_t^m,\overline{Z}_t^m)&\notag\\
\geq&\frac{\theta}{\psi}\delta^\psi\operatorname{exp}\left(-\frac{\psi}{\theta}\left(\overline{Y}_t^m-\int_0^t\left(\frac{1-\gamma}{2\gamma}\mu_s'\Sigma_s^{-1}\mu_s+(1-\gamma)(r_{s}\wedge m)-\delta \theta\right)ds\right)\right)\\
\geq&\frac{\theta}{\psi}\delta^\psi\operatorname{exp}\left(-\frac{\psi}{\theta}C_1T\right).
\end{align*}
Hence,  we consider the following BSDE
$${\underline{Y}_t^m}=\xi^m+\frac{\theta}{\psi}\delta^\psi\operatorname{exp}\left(-\frac{\psi}{\theta}C_1T\right)(T-t)-\int_t^T{\underline{Z}_s^m}d\hat{W}_s,$$
and it has an explicit form ${\underline{Y}_t^m}=\hat{\mathbf{E}}_t(\xi^m)++\frac{\theta}{\psi}\delta^\psi\operatorname{exp}\left(-\frac{\psi}{\theta}C_1T\right)(T-t).$
It follows from the comparison theorem (\cite{K00}, Theorem 2.6), dominance convergence theorem and Assumption \ref{IC} that
\begin{align*}
{\underline{Y}_t}\leq{\underline{Y}_t^m}\leq\mathcal{Y}_t^m\leq{\overline{Y}_t^m}\leq C_1T,
\end{align*}
where
\begin{align*}
\underline{Y}_t=\hat{\mathbf{E}}_t(\xi)+\frac{\theta}{\psi}\delta^\psi\operatorname{exp}\left(-\frac{\psi}{\theta}C_1T\right)(T-t).
\end{align*}

Using the localization technique introduced by \cite{BH06}, it implies the lower and upper bounds
\begin{align}\label{tildebound}
&\underline{Y}_t\leq\mathcal{Y}_t\leq\frac{1}{5}\ln\hat{\mathbf{E}}_t\left[\operatorname{exp}\left(5\xi+10(C_0+\gamma(\gamma-1)C_p)T\right)\right]-2(C_0+\gamma(\gamma-1)C_p)t.
\end{align}
Since $\mathcal{Y}_T=\xi$, subtracting $\int_0^\cdot\left(\frac{1-\gamma}{2\gamma}\mu_s'\Sigma_s^{-1}\mu_s+(1-\gamma)r_{s}-\delta \theta\right)ds$ on both sides of (\ref{tildebound}), then
\begin{align*}
&\mathcal{Y}_t-\int_0^t\left(\frac{1-\gamma}{2\gamma}\mu_s'\Sigma_s^{-1}\mu_s+(1-\gamma)r_{s}-\delta \theta\right)ds\\
\geq&\hat{\mathbf{E}}_t\left(\int_t^T\left(\frac{1-\gamma}{2\gamma}\mu_s'\Sigma_s^{-1}\mu_s+(1-\gamma)r_{s}-\delta \theta\right)ds\right)
+\frac{\theta}{\psi}\delta^\psi\operatorname{exp}\left(-\frac{\psi}{\theta}C_1T\right)(T-t)
\\\geq&(1-\gamma)\hat{\mathbf{E}}_t\left[\int_t^Tr_sds\right]+C_2(t)
\end{align*}
and
\begin{align*}
&\mathcal{Y}_t-\int_0^t\left(\frac{1-\gamma}{2\gamma}\mu_s'\Sigma_s^{-1}\mu_s+(1-\gamma)r_{s}-\delta \theta\right)ds\\
\leq&\frac{1}{5}\ln\hat{\mathbf{E}}_t\left[\operatorname{exp}\left(5\int_t^T\left((1-\gamma)r_{\operatorname{min}}-\delta \theta\right)ds+10(C_0+\gamma(\gamma-1)C_p)T\right)\right]-2(C_0+\gamma(\gamma-1)C_p)t\\
\leq& C_1T.
\end{align*}
Therefore, $$(1-\gamma)\hat{\mathbf{E}}_t\left[\int_t^Tr_sds\right]+C_2(t)\leq Y_t\leq C_1T,  \qquad \forall t\in[0,T].$$

As for the square integrability of $Z$, taking a stopping time sequence $\tau_n\to T$, we can obtain that
\begin{align*}
&\hat{\mathbf{E}}\left(\int_0^{\tau_n}Z_s\big(\frac{1}{2}I_k+\frac{1-\gamma}{2\gamma}\rho_s'\sigma_s'\Sigma_s^{-1}\sigma_s\rho_s\big)Z'_sds\right)\\
&\leq\mathcal{Y}_0-\hat{\mathbf{E}}\left(\mathcal{Y}_{\tau_n}\right)-\hat{\mathbf{E}}\left[\int_0^{\tau_n}\frac{\theta}{\psi}\delta^\psi e^{-\frac{\psi}{\theta}\mathcal{Y}_s}\operatorname{exp}\left({\frac{\psi}{\theta}\int_0^s\left(\frac{1-\gamma}{2\gamma}\mu_u'\Sigma_u^{-1}\mu_u+(1-\gamma) r_{u}-\delta \theta\right)du}\right)ds\right].
\end{align*}
Applying (\ref{zi}), the upper bound and the lower estimate of $\mathcal{Y}$ in ($\ref{tildebound}$), when $n\to+\infty$ yields
$$\hat{\mathbf{E}}\left(\int_0^T|Z_s|^2ds\right)<+\infty.$$
$\hfill\Box$

\subsection{Proof of Theorem \ref{g}}

This proof is similar to Proposition 2.2 and Lemma B.1 in \cite{X17}. Recall the definition of Epstein-Zin utility from (\ref{EZuc}),  for each $(c, p)\in\mathcal{S}_a$,
$$V_{\cdot}^c+\int_0^{\cdot}f(c_s,V_s^c)ds$$
is a local martingale by Proposition 2.2 in \cite{X17}.  Define $Y_t^c:=(1-\gamma)e^{-\delta\theta t}V_t^c$, using It\^{o}'s formula it implies that
\begin{align}\label{m1}
Y_\cdot^c+\int_0^\cdot F(c_s,Y_s^c)ds
\end{align}
is also a local martingale, where $F(c_t,y)=\delta\theta e^{-\delta t}c_t^{1-\frac{1}{\psi}}y^{1-\frac{1}{\theta}}$ is decreasing with $y$.

Let $\overline{V}^c:=\frac{\mathcal{W}^{1-\gamma}}{1-\gamma}e^{Y}$. From \eqref{mar},  it implies that $G_\cdot^{c,p}=\overline{V}_\cdot^c+\int_0^\cdot f(c_s,\overline{V}_s^c)ds$ is a local supermartingale. We claim that
\begin{align}\label{m2}
\overline{Y}_\cdot^c+\int_0^\cdot F(c_s,\overline{Y}_s^c)ds
\end{align}
is a local submartingale, where $\overline{Y}_t^c=(1-\gamma)e^{-\delta\theta t}\overline{V}_t^c$. Indeed, taking an appropriate stopping time sequence $\{\tau_n, n\geq 1\}$ with $\tau_{n}\stackrel{a.s.}\longrightarrow T$,  we obtain
$$\overline{V}_t^c+\int_0^tf(c_s,\overline{V}_s^c)ds\geq\mathbb{E}_t\left[\int_0^{\tau_n}f(c_s,\overline{V}_s^c)ds+\overline{V}_{\tau_n}^c\right],\quad t\in[0,\tau_n].$$
There exists an increasing process $A$ and a local square integrable process $\hat{Z}$ determined by  the Doob-Meyer decomposition and the martingale representation theorem (Theorem 16 in Chapter III, and Theorem 43 in Chapter IV from \cite{Pro05}), such that
$$\overline{V}_t^c+\int_0^tf(c_s,\overline{V}_s^c)ds+A_t=\overline{V}_0^c+\int_0^t\hat{Z}_{s}dB_s,  \qquad t\in[0,\tau_{n}].$$
It implies the following BSDE
$$\overline{V}_t^c=\overline{V}_{\tau_n}^c+\int_t^{\tau_n}f(c_s,\overline{V}_s^c)ds+A_{\tau_n}-A_t-\int_t^{\tau_n}\widehat{Z}_sdB_s,\quad t\in[0,\tau_n].$$
Applying It$\rm\hat{o}$'s formula to $(1-\gamma)e^{-\delta\theta t}\overline{V}_t^c$ yields
$$\overline{Y}_t^c=\overline{Y}_{\tau_n}^c+\int_t^{\tau_n}F(c_s,\overline{Y}_s^c)ds+(1-\gamma)e^{-\delta\theta t}(A_{\tau_n}-A_t)-\int_t^{\tau_n}\widehat{Z}_sdB_s,\quad t\in[0,\tau_n].$$
Therefore,
$$\overline{Y}_t^c+\int_0^tF(c_s,\overline{Y}_s^c)ds\leq\mathbb{E}_t\left[\int_0^{\tau_n}f(c_s,\overline{Y}_s^c)ds+\overline{Y}_{\tau_n}^c\right],\quad t\in[0,\tau_n].$$
Hence, the process (\ref{m2}) is a local submartingale.

Inspired by the third step of the proof of Proposition 2.2 in \cite{X17}, define
\[\alpha_t:=
\begin{cases}
\frac{F(c_t,\overline{Y}_t^c)-F(c_t,Y_t^c)}{\overline{Y}_t^c-Y_t^c}\quad\rm{for}\ \overline{Y}_t^c\neq Y_t^c,\\
0\quad\rm{for}\ \overline{Y}_t^c= Y_t^c.
\end{cases}
\]
Since $y\mapsto F(\cdot,y)$ is nonincreasing, $\alpha\leq0$. Comparing (\ref{m1}) and (\ref{m2}) we have
$$\overline{Y}_t^c-Y_t^c+\int_0^tF(c_s,\overline{Y}_s^c)-F(c_s,Y_s^c)ds=\overline{Y}_t^c-Y_t^c+\int_0^t\alpha_s\left(\overline{Y}_s^c-Y_s^c\right)ds.$$
It follows from It$\rm\hat{o}$'s formula and the class D property of $\overline{Y}^c$ and $Y^c$ that $e^{\int_0^\cdot\alpha_sds}\left(\overline{Y}_\cdot^c-Y_\cdot^c\right)$ is a submartingale satisfying
$$\mathbb{E}_t\left[e^{\int_0^T\alpha_sds}\left(\overline{Y}_T^c-Y_T^c\right)\right]\geq e^{\int_0^t\alpha_sds}\left(\overline{Y}_t^c-Y_t^c\right).$$
Note that $\overline{Y}_T^c=Y_T^c$, we finally get
\begin{align*}
V_t^c\leq\frac{\mathcal{W}_t^{1-\gamma}}{1-\gamma}e^{Y_t}.
\end{align*}

Moreover, since BSDE (\ref{mar}) has at least one solution, recall from (\ref{upo}) that
\begin{align}\label{up}
dG_t^{c,p}=&\frac{\mathcal{W}_t^{1-\gamma}}{1-\gamma}e^{Y_t}\Bigg[\left(Z_t+(1-\gamma)p_t'\rho_t\right)dW_t+(1-\gamma)p_t'\rho_t^\perp dW_t^\perp\notag\\
&+\big(\delta\theta e^{-\frac{Y_t}{\theta}}\hat{c}_t^{1-\frac{1}{\psi}}-(1-\gamma)\hat{c}_t+(1-\gamma)p_t'\sigma_t'\Sigma_t^{-1}(\mu_t+\sigma_t\rho_t Z_t)\notag\\
&-\frac{\gamma(1-\gamma)}{2})p_t'p_t-H(\cdot,Y_t,Z_t)
+(1-\gamma)r_t-\delta\theta+\frac{1}{2}Z_tZ_t'\big)dt\Bigg].
\end{align}
If $(c^*,p^*)\in\mathcal{S}_a$, then $\overline{V}^{c^*}+\int_0^\cdot f(c_s^*,\overline{V}_s^{c^*})ds$ is a local martingale due to
(\ref{generator-complete}) and (\ref{up}). Taking a  localizing sequence $(\sigma_n)_{n\geq1}$, we have
$$\overline{V}_0^{c^*}+\delta\theta\mathbf{E}\left[\int_0^{T\wedge\sigma_n}\overline{V}_s^{c^*}ds\right]=\mathbf{E}\left[\overline{V}_{T\wedge\sigma_n}^{c^*}+\int_0^{T\wedge\sigma_n}\delta\frac{\left(c_s^*\right)^{1-\frac{1}{\psi}}}{1-\frac{1}{\psi}}\left((1-\gamma)\overline{V}_s^{c^*}\right)^{1-\frac{1}{\theta}}ds\right].$$
Since $\overline{V}^{c^*}\leq0$ and $\psi>1$, the integrand on the left side of the equality is nonpositive and the integrand on the right side is nonnegative. Therefore, the class D property of $\overline{V}^{c^*}$ and the monotone convergence theorem imply that
\begin{align*}
\mathbf{E}\left[\int_0^Tf\left(c_s^*,\frac{\left(\mathcal{W}_s^*\right)^{1-\gamma}}{1-\gamma}e^{Y_s}\right)ds+\frac{\left(\mathcal{W}_T^*\right)^{1-\gamma}}{1-\gamma}\right]=\frac{\omega^{1-\gamma}}{1-\gamma}e^{Y_0}.
\end{align*}
In this way, we confirm that $\frac{\omega^{1-\gamma}}{1-\gamma}e^{Y_0}$ is the optimal utility.  Then the corresponding strategy $(c^*,p^*)$ is the optimal strategy for the problem (\ref{max}).
$\hfill\Box$

\subsection{Proof of Theorem \ref{main}}
According to the definition of $W^\rho$, we have
\begin{align*}
&(1-\gamma)(p^*)'dW^\rho+ZdW\\
=&\left((1-\gamma)(p^*)'\rho+Z\right)dW+(1-\gamma)(p^*)'\rho^\perp dW^\perp\\
=:&M^{(1)}dW+M^{(2)}dW^\perp.
\end{align*}
Bear in mind that the randomness of market parameters comes only from $W$, so we start with $\mathscr{E}\left(\int M^{(1)}dW\right)$. Define a stopping time sequence $\tau_n:=\inf\left\{t\geq0: X_t\notin E_n\right\}\wedge T$. Now we claim that $Y_{\cdot\wedge\tau_n}$ is bounded. Combined with Proposition \ref{Yestimate}, it is sufficient to prove that
$\hat{\mathbf{E}}_{\cdot\wedge\tau_n}\left[\int_{\cdot\wedge\tau_n}^{T}r_s d s\right]$ is bounded from above. Similar to \cite{X17}, using Theorem 1 of \cite{HS00}, the Feynman-Kac formula guarantees that
$$y(t, x):=\hat{\mathbf{E}}\left(\int_t^Tr_sds\Big|X_t=x\right),\quad t\in[0,T], x\in E$$
is in $C^{1,2}([0,T]\times E)$ and $y$ is the unique solution to the following PDE
\[
\begin{dcases}
&\partial_ty+\mathscr{L}y+r=0,\\
&y(T,x)=0,
\end{dcases}
\]
where $\mathscr{L}$ is the infinitesimal generator of $X$ under $\hat{\mathbb{P}}$. Hence $y(\cdot\wedge\tau_n,X_{\cdot\wedge\tau_n})=\hat{\mathbf{E}}_{\cdot\wedge\tau_n}\left[\int_{\cdot\wedge\tau_n}^{T}r_sd s\right]$ is bounded due to the compactness of $\overline{E}_n$. Consider the following truncated BSDE,
$$Y_t=Y_{\tau_n}+\int_t^{\tau_n}H(s,Y_s,Z_s)ds-\int_t^{\tau_n}Z_sdW_s,\quad t\in[0,\tau_n].$$
By Assumption \ref{para}, $Y_{\cdot\wedge\tau_n}$ is then bounded. Since $H$ has quadratic growth in $z$, then Lemma 3.1 in \cite{M09} implies that $\int_0^{\cdot\wedge\tau_n}Z_sdW_s\in\operatorname{BMO}(\mathbb{P})$. Observe that there exist two constants $k_1(n)$ and $k_2(n)$ depending on $n$, such that for $t\in[0,\tau_n]$,
\begin{align}\label{pz}
|p^*_{t}|&\leq\left|p^*_{t}-\frac{1}{\gamma}\sigma_{t}'\Sigma_{t}^{-1}(\mu_{t}+\sigma_{t}\rho_{t}Z'_t)\right|+\left|\frac{1}{\gamma}\sigma_{t}'\Sigma_{t}^{-1}(\mu_{t}+\sigma_{t}\rho_{t}Z'_t)\right|\notag\\
&=\operatorname{min}_{p_t\in\mathcal{A}_t}\left|p_{t}-\frac{1}{\gamma}\sigma_{t}'\Sigma_{t}^{-1}(\mu_{t}+\sigma_{t}\rho_{t}Z'_t)\right|+\left|\frac{1}{\gamma}\sigma_{t}'\Sigma_{t}^{-1}(\mu_{t}+\sigma_{t}\rho_{t}Z'_t)\right|\notag\\
&\leq\operatorname{min}_{p_t\in\mathcal{A}_t}|p_t|+2\left|\frac{1}{\gamma}\sigma_{t}'\Sigma_{t}^{-1}(\mu_{t}+\sigma_{t}\rho_{t}Z'_t)\right|\\
&\leq k_1(n)+k_2(n)|Z_t|. \notag
\end{align}
The last inequality follows from (\ref{bap}). This means that $\int_0^{\cdot\wedge\tau_n}p^*_sdW_s\in\operatorname{BMO}(\mathbb{P})$. Therefore, $\mathscr{E}\left(\int M^{(1)}dW\right)_{\cdot\wedge\tau_n}$ is a uniformly integrable $\mathbb{P}$-martingale by Theorem 2.3 in \cite{K94}. Therefore,
\begin{align}\label{npm}
\frac{d\mathbb{P}^n}{d\mathbb{P}}:=\mathscr{E}\left(\int M^{(1)}dW\right)_{\tau_n}
\end{align}
defines a new probability measure $\mathbb{P}^n$ equivalent to $\mathbb{P}$ on $\mathscr{F}_{\tau_n}$. Then we characterize $\mathbb{P}^n$ with the following lemma.

\begin{lem}\label{zero}
Let Assumptions \ref{ccc}, \ref{para}, \ref{IC} and \ref{Lfo} hold. Then
$\lim_{n\to\infty}\mathbb{P}^n\left(\tau_n<T\right)=0$.
\end{lem}
\noindent{\bf{Proof.}}  ~We rewrite (\ref{mar}) as follows
$$Y_t=Y_0-\int_0^tH(s,Y_s,Z_s)ds+\int_0^tZ_sdW_s,\quad t\in[0,T].$$
Since $\phi\in C^{2}(E)$, applying It$\rm\hat{o}$'s formula to $\phi(X_t)$, it yields that
\begin{align*}
\phi(X_t)=\phi(x)+\int_0^tb_s'\nabla\phi+\frac{1}{2}\sum_{i,j=1}^{k}A_{ij}\partial^2_{x_ix_j}\phi ds+\int_0^t\nabla\phi'a_sdW_s.
\end{align*}
From (\ref{npm}), $W^n:=W-\int_0^\cdot\left[(1-\gamma)\left(p_s^*\right)'\rho_s+Z_s\right]ds$ is a Brownian motion on $[0,\tau_n]$ under $\mathbb{P}^n$, and we obtain that
\begin{align*}
&Y_t-\phi(X_t)\\
=&Y_0-\phi(x)+\int_0^t-H(s,Y_s,Z_s)-b_s'\nabla\phi-\frac{1}{2}\sum_{i,j=1}^{k}A_{ij}\partial^2_{x_ix_j}\phi ds+\int_0^tZ_s-\nabla\phi'a_sdW_s\\
=&Y_0-\phi(x)+\int_0^t\frac{\gamma(1-\gamma)}{2}\operatorname{dist}^2\left(\mathcal{A}_s, \frac{1}{\gamma}\sigma_s'\Sigma_s^{-1}(\mu_s+\sigma_s\rho_sZ_s')\right)\\
&-Z_s\left(\frac{1}{2}I_k+\frac{1-\gamma}{2\gamma}\rho_s'\sigma_s'\Sigma_s^{-1}\sigma_s\rho_s\right)Z_s'-\frac{1-\gamma}{\gamma}\mu_s'\Sigma_s^{-1}\sigma_s\rho_sZ_s'-\frac{\theta}{\psi}\delta^\psi e^{-\frac{\psi}{\theta}Y_s}\\
&-\frac{1-\gamma}{2\gamma}\mu_s'\Sigma_s^{-1}\mu_s-(1-\gamma) r_{s}+\delta \theta-b_s'\nabla\phi-\frac{1}{2}\sum_{i,j=1}^{k}A_{ij}\partial^2_{x_ix_j}\phi ds\\
&+\int_0^t\left(Z_s-\nabla\phi'a_s\right)\left(dW_s^n+\left[(1-\gamma)\left(p_s^*\right)'\rho_s+Z_s\right]ds\right)\\
=&Y_0-\phi(x)+\int_0^t\frac{\gamma(1-\gamma)}{2}\operatorname{dist}^2\left(\mathcal{A}_s, \frac{1}{\gamma}\sigma_s'\Sigma_s^{-1}(\mu_s+\sigma_s\rho_sZ_s')\right)\\
&+Z_s\left(\frac{1}{2}I_k-\frac{1-\gamma}{2\gamma}\rho_s'\sigma_s'\Sigma_s^{-1}\sigma_s\rho_s\right)Z_s'+\left[(1-\gamma)\left(p_s^*\right)'\rho_s-\frac{1-\gamma}{\gamma}\mu_s'\Sigma_s^{-1}\sigma_s\rho_s-\nabla\phi'a_s\right]Z_s'\\
&-(1-\gamma)\left(p_s^*\right)'\rho_sa_s'\nabla\phi-\frac{\theta}{\psi}\delta^\psi e^{-\frac{\psi}{\theta}Y_s}-\frac{1-\gamma}{2\gamma}\mu_s'\Sigma_s^{-1}\mu_s-(1-\gamma) r_{s}+\delta \theta-b_s'\nabla\phi\\
&-\frac{1}{2}\sum_{i,j=1}^{k}A_{ij}\partial^2_{x_ix_j}\phi ds+\int_0^t\left(Z_s-\nabla\phi'a_s\right)dW_s^n.
\end{align*}

Different from the proof of Lemma B.2 in \cite{X17}, since there is no explicit solution for $p^*$ due to closed constraints in our situation,  we will perform more refined analysis and estimations for the drift term in the following.

For any $s\in[0,\tau_{n}]$, it implies from (\ref{bap}) that
\begin{align*}
&\frac{\gamma(1-\gamma)}{2}\operatorname{dist}^2\left(\mathcal{A}_s, \frac{1}{\gamma}\sigma_s'\Sigma_s^{-1}(\mu_s+\sigma_s\rho_sZ_s')\right)\\
=&\frac{\gamma(1-\gamma)}{2}\min_{p_s\in\mathcal{A}_s}\left|p_s-\frac{1}{\gamma}\sigma_s'\Sigma_s^{-1}(\mu_s+\sigma_s\rho_sZ_s')\right|^2\\
\geq&\frac{1-\gamma}{2\gamma}\Big|\sigma_s'\Sigma_s^{-1}(\mu_s+\sigma_s\rho_sZ_s')\Big|^2\\
=&\frac{1-\gamma}{2\gamma}Z_s\rho_s'\sigma_s'\Sigma_s^{-1}\sigma_s\rho_sZ_s'+\frac{1-\gamma}{\gamma}\mu'_s\Sigma^{-1}_s\sigma_s\rho_sZ_s'+\frac{1-\gamma}{2\gamma}\mu'_s\Sigma^{-1}_s\mu_s.
\end{align*}
Therefore,
\begin{align}\label{ulf}
&Y_t-\phi(X_t)\notag\\
\geq&Y_0-\phi(x)+\int_0^t\frac{1}{2}Z_sZ_s'+\left[(1-\gamma)\left(p_s^*\right)'\rho_s-\nabla\phi'a_s\right]Z_s'-(1-\gamma)\left(p_s^*\right)'\rho_sa_s'\nabla\phi\notag\\
&-\frac{\theta}{\psi}\delta^\psi e^{-\frac{\psi}{\theta}Y_s}-(1-\gamma) r_{s}+\delta \theta-b_s'\nabla\phi-\frac{1}{2}\sum_{i,j=1}^{k}A_{ij}\partial^2_{x_ix_j}\phi ds+\int_0^t\left(Z_s-\nabla\phi'a_s\right)dW_s^n.
\end{align}
Based on the inequality \eqref{pz}, for $t\in[0,T]$, we get
\begin{align*}
|p^*_t|
&\leq\min_{p_t\in\mathcal{A}_t}\left|p_t\right|+2\left|\frac{1}{\gamma}\sigma_t'\Sigma_t^{-1}(\mu_t+\sigma_t\rho_tZ'_t)\right|\leq C_p+\frac{2}{\gamma}\left|\sigma_t'\Sigma_t^{-1}\mu_t\right|+\frac{2}{\gamma}\left|\sigma_t'\Sigma_t^{-1}\sigma_t\rho_tZ'_t\right|.
\end{align*}
Then from algebraic inequalities, it yields the following estimations
\begin{align}\label{e1}
(p_t^*)'\rho_tZ_t'&\leq\left|(p_t^*)'\rho_tZ_t'\right|\leq\left|p_t^*\right|\left|\rho_t\right|\left|Z_t\right|\notag\\
&\leq C_p\left|\rho_t\right|\left|Z_t\right|+\frac{2}{\gamma}\left|\sigma_t'\Sigma_t^{-1}\mu_t\right|\left|\rho_t\right|\left|Z_t\right|+\frac{2}{\gamma}\left|\sigma_t'\Sigma_t^{-1}\sigma_t\rho_t\right|\left|\rho_t\right|\left|Z_t\right|^2
\end{align}
and
\begin{align}\label{e2}
(p_t^*)'\rho_ta_t'\nabla\phi&\geq-\left|(p_t^*)'\rho_ta_t'\nabla\phi\right|\geq-\left|p_t^*\right|\left|\rho_ta_t'\nabla\phi\right|\notag\\
&\geq-C_p\left|\rho_ta_t'\nabla\phi\right|-\frac{2}{\gamma}\left|\sigma_t'\Sigma_t^{-1}\mu_t\right|\left|\rho_ta_t'\nabla\phi\right|-\frac{2}{\gamma}\left|\sigma_t'\Sigma_t^{-1}\sigma_t\rho_tZ'_t\right|\left|\rho_ta_t'\nabla\phi\right|.
\end{align}
Plugging (\ref{e1}) and (\ref{e2}) into (\ref{ulf}), it derives that
\begin{align*}
&\frac{1}{2}Z_tZ_t'+\left[(1-\gamma)\left(p_t^*\right)'\rho_t-\nabla\phi'a_t\right]Z_t'-(1-\gamma)\left(p_t^*\right)'\rho_ta_t'\nabla\phi\\
\geq&\frac{1}{2}Z_tZ_t'+(1-\gamma)C_p|\rho_t||Z_t|+\frac{2(1-\gamma)}{\gamma}\left|\sigma_t'\Sigma_t^{-1}\mu_t\right|\left|\rho_t\right|\left|Z_t\right|+\frac{2(1-\gamma)}{\gamma}\left|\sigma_t'\Sigma_t^{-1}\sigma_t\rho_t\right|\left|\rho_t\right|\left|Z_t\right|^2\\
&-\left|\nabla\phi'a_t\right|\left|Z_t\right|+(1-\gamma)C_p\left|\rho_ta_t'\nabla\phi\right|+\frac{2(1-\gamma)}{\gamma}\left|\sigma_t'\Sigma_t^{-1}\mu_t\right|\left|\rho_ta_t'\nabla\phi\right|\\
&+\frac{2(1-\gamma)}{\gamma}\left|\sigma_t'\Sigma_t^{-1}\sigma_t\rho_t\right|\left|\rho_ta_t'\nabla\phi\right|\left|Z_t\right|\\
=&\left(\frac{1}{2}+\frac{2(1-\gamma)}{\gamma}\left|\sigma_t'\Sigma_t^{-1}\sigma_t\rho_t\right|\left|\rho_t\right|\right)\left|Z_t\right|^2+\bigg((1-\gamma)C_p|\rho_t|+\frac{2(1-\gamma)}{\gamma}\left|\sigma_t'\Sigma_t^{-1}\mu_t\right|\left|\rho_t\right|-\left|\nabla\phi'a_t\right|\\
&+\frac{2(1-\gamma)}{\gamma}\left|\sigma_t'\Sigma_t^{-1}\sigma_t\rho_t\right|\left|\rho_ta_t'\nabla\phi\right|\bigg)\left|Z_t\right|+\frac{2(1-\gamma)}{\gamma}\left|\sigma_t'\Sigma_t^{-1}\mu_t\right|\left|\rho_ta_t'\nabla\phi\right|+(1-\gamma)C_p\left|\rho_ta_t'\nabla\phi\right|.
\end{align*}

Therefore, we can get
\begin{align*}
&Y_t-\phi(X_t)\\
\geq&Y_0-\phi(x)+\int_0^t\left(\frac{1}{2}+\frac{2(1-\gamma)}{\gamma}\left|\sigma_s'\Sigma_s^{-1}\sigma_s\rho_s\right|\left|\rho_s\right|\right)\Bigg(|Z_s|\\
&+\frac{(1-\gamma)C_p|\rho_s|+\frac{2(1-\gamma)}{\gamma}\left|\sigma_s'\Sigma_s^{-1}\mu_s\right|\left|\rho_s\right|-\left|\nabla\phi'a_s\right|+\frac{2(1-\gamma)}{\gamma}\left|\sigma_s'\Sigma_s^{-1}\sigma_s\rho_s\right|\left|\rho_sa_s'\nabla\phi\right|}{1+\frac{4(1-\gamma)}{\gamma}\left|\sigma_s'\Sigma_s^{-1}\sigma_s\rho_s\right|\left|\rho_s\right|}\Bigg)^2\\
&-\frac{\theta}{\psi}\delta^\psi e^{-\frac{\psi}{\theta}Y_s}-(1-\gamma) r_{s}+\delta \theta-b_s'\nabla\phi-\frac{1}{2}\sum_{i,j=1}^{k}A_{ij}\partial^2_{x_ix_j}\phi\\
&+\frac{2(1-\gamma)}{\gamma}\left|\sigma'_s\Sigma_s^{-1}\mu_s\right|\left|\rho_sa_s'\nabla\phi\right|+(1-\gamma)C_p\left|\rho_sa_s'\nabla\phi\right|\\
&-\frac{\left((1-\gamma)C_p|\rho_s|+\frac{2(1-\gamma)}{\gamma}\left|\sigma_s'\Sigma_s^{-1}\mu_s\right|\left|\rho_s\right|-\left|\nabla\phi'a_s\right|+\frac{2(1-\gamma)}{\gamma}\left|\sigma_s'\Sigma_s^{-1}\sigma_s\rho_s\right|\left|\rho_sa_s'\nabla\phi\right|\right)^2}{2+\frac{8(1-\gamma)}{\gamma}\left|\sigma_s'\Sigma_s^{-1}\sigma_s\rho_s\right|\left|\rho_s\right|} ds\\
&+\int_0^t\left(Z_s-\nabla\phi'a_s\right)dW_s^n\\
\geq&Y_0-\phi(x)+\int_0^t\left(-\frac{\theta}{\psi}\delta^\psi e^{-\frac{\psi}{\theta}Y_s}-(1-\gamma) r_{s}+\delta \theta-\mathfrak{F}\left(\phi\right)\right)ds+\int_0^t\left(Z_s-\nabla\phi'a_s\right)dW_s^n.
\end{align*}
The last inequality derives from the definition of $\mathfrak{F}$ and the fact that $\frac{1}{2}+\frac{2(1-\gamma)}{\gamma}\left|\sigma'\Sigma^{-1}\sigma\rho\right|\left|\rho\right|>0$. Since $\theta<0$, $r$ is bounded from below, and $\mathfrak{F}$ is bounded from above, then we can find a negative constant $C_3$, such that,
\begin{align}\label{con}
Y_{\tau_n}-\phi(X_{\tau_n})\geq Y_0-\phi(x)+C_3\tau_n+\int_0^{\tau_n}\left(Z_s-\nabla\phi'a_s\right)dW_s^n.
\end{align}
Then by Theorem 3.6 of \cite{K94}, $\int_0^{\cdot\wedge\tau_n}\left(Z_s-\nabla\phi'a_s\right)dW_s^n\in \operatorname{BMO}(\mathbb{P}^n)$. Taking expectation under $\mathbb{P}^n$ on both sides of (\ref{con}), we obtain
\begin{align}\label{Ei}
\mathbf{E}^{\mathbb{P}^n}\left[Y_{\tau_n}-\phi(X_{\tau_n})\right]\geq Y_0-\phi(x)+C_3T>-\infty,\qquad\forall n.
\end{align}
On the other hand, by Proposition  \ref{Yestimate} and  the definition of $\tau_{n}$, it derives
\begin{align*}
Y_{\tau_n}-\phi(X_{\tau_n})=& Y_{\tau_n}-\phi(X_{\tau_n})\mathbb{I}_{\{\tau_n=T\}}-\phi(X_{\tau_n})\mathbb{I}_{\{\tau_n<T\}}\\
\leq& C_{1}T-\operatorname{inf}_{x\in E_n}\phi(x)\mathbb{I}_{\{\tau_n=T\}}-\operatorname{inf}_{x\in \partial E_n}\phi(x)\mathbb{I}_{\{\tau_n<T\}}.
\end{align*}
By Assumption \ref{Lfo}(i), there exists a constant $C_4$ such that $\operatorname{inf}_{x\in E}\phi(x)\geq C_{4}$. Thus the previous inequality implies that
\begin{align*}
\mathbf{E}^{\mathbb{P}^n}\left[Y_{\tau_n}-\phi(X_{\tau_n})\right]\leq C_{1}T+|C_4|-\operatorname{inf}_{x\in \partial E_n}\phi(x)\mathbb{P}^n(\tau_n<T).
\end{align*}
From inequality (\ref{Ei}), Assumption \ref{Lfo}(\romannumeral1) derives that $\mathbb{P}^n(\tau_n<T)\to0$.
$\hfill\Box$\\[-10pt]

Using Lemma \ref{zero}, we can verify the martingale property of (\ref{mse}).
\begin{lem}\label{mp}
Let Assumptions \ref{ccc}, \ref{para}, \ref{IC} and \ref{Lfo} hold. Then $M$ is a $\mathbb{P}$-martingale.
\end{lem}

\noindent\textbf{Proof.}
By Lemma \ref{zero}, we have
\begin{align*}
&\mathbf{E}\left[\mathscr{E}\left(\int M^{(1)}_sdW_s\right)_T\right]\\
=&\lim_{n\to\infty}\mathbf{E}\left[\mathscr{E}\left(\int M^{(1)}_sdW_s\right)_{\tau_n}\right]-\lim_{n\to\infty}\mathbf{E}\left[\mathscr{E}\left(\int M^{(1)}_sdW_s\right)_{\tau_n}\mathbb{I}_{\{\tau_n<T\}}\right]\\
=&1-\lim_{n\to\infty}\mathbb{P}^n\left(\tau_n<T\right)=1.
\end{align*}
This implies the martingale property of $\mathscr{E}\left(\int M_s^{(1)}dW_{s}\right)$ on $[0,T]$. Then we can confirm that $M$ is a $\mathbb{P}$-martingale. In fact, letting $\mathcal{F}^W=\sigma\left(W_s, s\in[0,T]\right)$, since $M^{(2)}$ is independent of $W^\perp$, Lemma 4.8 of \cite{KK07} implies that for any $t\in[0,T]$,
\begin{align*}
\mathbf{E}\left(M_t\right)
&=\mathbf{E}\left[\mathscr{E}\left(\int M_s^{(1)}dW_s+\int M_s^{(2)}dW_s^\perp\right)_t\right]\\
&=\mathbf{E}\left[\mathscr{E}\left(\int M_s^{(1)}dW_s\right)_t\mathbf{E}\left[\mathscr{E}\left(\int M_s^{(2)}dW_s^\perp\right)_t\big|\mathcal{F}^W\right]\right]\\
&=\mathbf{E}\left[\mathscr{E}\left(\int M_s^{(1)}dW_s\right)_t\right]\\
&=1.
\end{align*}$\hfill\Box$


\noindent\textbf{Proof of Theorem \ref{main}.} By Theorem \ref{g}, we only need to show $(c^{*}, p^{*})\in \mathcal{S}_{a}$.

 Firstly,  $\left(\mathcal{W}^*\right)^{1-\gamma}e^Y$ is of class D.   In fact, applying It$\rm\hat{o}$'s formula to $\left(\mathcal{W}^*\right)^{1-\gamma}e^Y$,  we have
\begin{align}\label{UI}
\left(\mathcal{W}_t^*\right)^{1-\gamma}e^{Y_t}=\omega^{1-\gamma}e^{Y_0}\operatorname{exp}\left(\int_0^t\left(\theta\delta-\theta\delta^\psi e^{-\frac{\psi}{\theta}Y_s}\right)ds\right)M_t,
\end{align}
where $M$ is defined in (\ref{mse}). By Lemma \ref{mp}, $M$ is of class D and other terms are bounded. Then $\left(\mathcal{W}^*\right)^{1-\gamma}e^Y$ is of class D on $[0,T]$.

Next, we claim $c^{*}\in\mathcal{C}_{a}$.  Since $Y_{T}=0$ and  $\left(\mathcal{W}^*\right)^{1-\gamma}e^Y$ is of class D, then $\mathbf{E}\big[(\mathcal{W}_{T}^*)^{1-\gamma}\big]<\infty$. Moreover,  $e^{-\delta t}(c_t^*)^{1-\frac{1}{\psi}}=\delta^{\psi-1}e^{-\delta t-\frac{\psi-1}{\theta}Y_t}\left(\mathcal{W}_t^*\right)^{1-\frac{1}{\psi}}$, it is only to show $\mathbf{E}\big[\int_0^T(\mathcal{W}_s^*)^{1-\frac{1}{\psi}}ds\big]<\infty.$

Indeed, H\"older inequality and Assumption \ref{rnm} imply that
\begin{align*}
&\mathbf{E}\left[\int_0^T\left(\mathcal{W}_s^*\right)^{1-\frac{1}{\psi}}ds\right]\\
=&\int_0^T\mathbf{E}^0\left[\operatorname{exp}\left((1-\frac{1}{\psi})\int_0^sr_udu\right)\operatorname{exp}\left(-(1-\frac{1}{\psi})\int_0^sr_udu\right)\left(\mathcal{W}_s^*\right)^{1-\frac{1}{\psi}}\mathscr{E}\left(\int\lambda_udW_u^0\right)_T\right]ds\\
\leq&\int_0^T\mathbf{E}^0\left(\operatorname{exp}\left((1-\frac{1}{\psi})\int_0^Tr_u^+du\right)\mathscr{E}\left(\int\lambda_udW_u^0\right)_T\operatorname{exp}\left(-(1-\frac{1}{\psi})\int_0^sr_udu\right)\left(\mathcal{W}_s^*\right)^{1-\frac{1}{\psi}}\right)ds\\
\leq&\mathbf{E}^0\left(\operatorname{exp}\left((\psi-1)\int_0^Tr_u^+du\right)\mathscr{E}\left(\int\lambda_udW_u^0\right)_T^\psi\right)^{\frac{1}{\psi}}\int_0^T\mathbf{E}^0\left(\operatorname{exp}\left(-\int_0^sr_udu\right)\mathcal{W}_s^*\right)^{1-\frac{1}{\psi}}ds.
\end{align*}
The statement is then confirmed due to the supermartingale property of $\operatorname{exp}\left(-\int_0^\cdot r_udu\right)\mathcal{W}^*$ under $\mathbb{P}_0$ as well as Assumption \ref{rnm}.
$\hfill\Box$

\subsection{Proof of Proposition \ref{exp2}}

In order to prove Proposition \ref{exp2} and the sequent Proposition \ref{exp1}, we need a lemma about the Laplace transform of a square root process. One can refer to Equation 2.k of \cite{PY82} or Lemma C.1 of \cite{X17}.
\begin{lem}\label{LLT}
A square root process $X$ is given by
$$dX_t=(\alpha-\beta X_t)dt+a\sqrt{X_t}dW_t,$$
where $W$ is a 1-dimensional Brownian motion. If $\zeta<\frac{\beta^2}{2a^2}$,
then for any $T\geq0$, the Laplace transform
$$\mathbf{E}\left(e^{\zeta\int_0^TX_sds}\Big|X_0=x\right)$$
is well defined.
\end{lem}
Assumptions \ref{IC}, \ref{Lfo} and \ref{rnm} are all verified in what follows. Then Theorem \ref{main} holds.

Step 1. Verification of Assumption \ref{IC}. Note that $$\mathscr{E}\left(\int \frac{1-\gamma}{\gamma}\mu\Sigma^{-1}\sigma\rho dW_s\right)_T=\mathscr{E}\left(\int\frac{1-\gamma}{\gamma}(\lambda_0+\lambda_1((-100)\vee X_{s}\wedge100))\rho dW_s\right)_T.$$
Obviously, $\hat{\mathbb{P}}$ is well-defined.  One can refer to Eq.(3.17) of Chapter 5 of \cite{KS06} for the integrability of $X$ because $X$ is another Ornstein-Uhlenbeck process with modified linear drift under $\hat{\mathbb{P}}$. Therefore, Assumption \ref{IC} is satisfied.

Step 2. Verification of Assumption \ref{Lfo}. Considering $\phi(x)=c_0x^2$ for some fixed constant $c_0>0$, we can prove that $\phi(x)\nearrow+\infty$ as $|x|\nearrow+\infty$. Meanwhile the operator $\mathfrak{F}$ reads
\begin{align*}
\mathfrak{F}[\phi]=c_0\left(\frac{a^2\left(\frac{2(1-\gamma)\rho^2}{\gamma}-1\right)^2}{\frac{1}{2}+\frac{2(1-\gamma)\rho^2}{\gamma}}-2b\right)x^2+\text{lower order terms in x}.
\end{align*}
Then sufficiently small positive constants $c_0$ lead to $\mathfrak{F}[\phi](x)\searrow-\infty$ as $x\searrow0$ or $x\nearrow+\infty$ when Condition (\romannumeral3) in Proposition \ref{exp2} holds. Therefore $\mathfrak{F}[\phi]$ is bounded from above on $(0,+\infty)$.

Step 3. Verification of Assumption \ref{rnm}. In order to apply Lemma \ref{LLT}, we change the dynamics of $X$ to the following form
$$dX_t=-(b-(\psi-1)a\lambda_1\rho)X_tdt+ad\tilde{W}_t.$$
Let $Y:=X^2$, it thus has the dynamics
$$dY_t=(a^2-2(b-(\psi-1)a\lambda_1\rho)Y_t)dt+2a\sqrt{Y_t}d\tilde{W}_t,$$
where $\tilde{W}:=W_0+\int_0^\cdot(\psi-1)\lambda_0\rho ds$ is a $\tilde{\mathbb{Q}}$-Brownian motion and $\tilde{\mathbb{Q}}$ is defined as $$\frac{d\tilde{\mathbb{Q}}}{d\mathbb{P}_0}=\mathscr{E}\left(-(\psi-1)\lambda_0\rho W_T\right).$$
According to Remark \ref{degen}, for sufficiently small $\epsilon>0$, it remains to prove
\begin{align}\label{lastintegral}
\mathbf{E}^0\left[\operatorname{exp}\left((1+\epsilon)(\psi-1)\left(r_0^{+}T+r_1\int_0^T X_s^{+} ds\right)\right)\right]<+\infty.
\end{align}
In fact, H\"older inequality implies that for $p>0$,
\begin{align*}
&\mathbf{E}^0\left[\operatorname{exp}\left((1+\epsilon)(\psi-1)r_1\int_0^TX_s^{+}ds\right)\right]\\
\leq&\mathbf{E}^0\left[\operatorname{exp}\left((1+\epsilon)(\psi-1)r_1\left(\int_0^TX_s^2ds+\frac{1}{4}T\right)\right)\right]\\
=&\mathbf{E}^{\tilde{\mathbb{Q}}}\left[\frac{d\mathbb{P}_0}{d\tilde{\mathbb{Q}}}\operatorname{exp}\left((1+\epsilon)(\psi-1)r_1\left(\int_0^TX_s^2ds+\frac{1}{4}T\right)\right)\right]\\
\leq&\mathbf{E}^{\tilde{\mathbb{Q}}}\left[\left(\frac{d\mathbb{P}_0}{d\tilde{\mathbb{Q}}}\right)^{\frac{1+p}{p}}\right]^{\frac{p}{1+p}}\mathbf{E}^{\tilde{\mathbb{Q}}}\left[\operatorname{exp}\left((1+p)(1+\epsilon)(\psi-1)r_1\left(\int_0^TX_s^2ds+\frac{1}{4}T\right)\right)\right]^{\frac{1}{1+p}}.
\end{align*}
The first expectation on the right-hand side is finite because of the fact that $\frac{d\mathbb{P}_0}{d\tilde{\mathbb{Q}}}$ has any finite moment. As for the second expectation, Lemma \ref{LLT} leads to the inequality that for sufficiently small $p$ and $\epsilon$,
$$(\psi-1)r_1<\frac{4(b-(\psi-1)a\lambda_1\rho)^2}{8a^2}.$$
Note that this is Condition (\romannumeral4) in Proposition \ref{exp2}.
$\hfill\Box$

\subsection{Proof of Proposition \ref{exp1}}
It is clear that the statement of Theorem \ref{main} holds when Assumptions \ref{IC}, \ref{Lfo} and \ref{rnm} are all verified.\\[-12pt]

Step 1. Verification of Assumption \ref{IC}. Direct calculation indicates that $$\mathscr{E}\left(\frac{1-\gamma}{\gamma}\mu\Sigma^{-1}\sigma\rho dW_s\right)_T=\mathscr{E}\left(\frac{1-\gamma}{\gamma}\lambda\rho dW_s\right)_T.$$
Hence $\hat{\mathbb{P}}$ is well-defined and $\hat{W}=W-\int_0^\cdot\frac{1-\gamma}{\gamma}\lambda\rho ds$ is a Brownian motion under $\hat{\mathbb{P}}$. In addition, it follows from
$$dX_t=\left(bl+\frac{1-\gamma}{\gamma}a\rho\lambda\sqrt{X_t}-bX_t\right)dt+a\sqrt{X_t}d\hat{W}_t$$
that $X_t-X_0\leq blt+\int_0^ta\sqrt{X_t}d\hat{W_t}$. Then we have $\hat{\mathbf{E}}\left(X_t-X_0\right)\leq blt$. Hence
$$\hat{\mathbf{E}}\left[\int_0^Tr_sds\right]=r_0T+r_1\int_0^T\hat{\mathbf{E}}[X_s]ds<+\infty.$$

Step 2. Verification of Assumption \ref{Lfo}. It is easy to check that $\phi(x)\nearrow+\infty$ as $x\nearrow+\infty$ or $x\searrow0$ when taking $\phi(x)=c_1x-c_2\ln x$, where $c_1$ and $c_2$ are both sufficiently small positive constants. Moreover, the operator $\mathfrak{F}$ can be written as
\begin{align*}
\mathfrak{F}[\phi]=&blc_1+bc_2+\frac{\frac{2(1-\gamma)^2}{\gamma^2}\lambda^2\rho^2}{1+\frac{4(1-\gamma)}{\gamma}\rho^2}-\frac{a^2c_1c_2\left(\frac{2(1-\gamma)}{\gamma}\rho^2-1\right)^2}{1+\frac{4(1-\gamma)}{\gamma}\rho^2}+\frac{(1-\gamma)^2\rho^2C_p\left(C_p+\frac{4\lambda}{\gamma}\right)}{2\left(1+\frac{4(1-\gamma)}{\gamma}\rho^2\right)}\\
&+c_1\left[\frac{\left(\frac{2(1-\gamma)}{\gamma}\rho^2-1\right)^2a^2c_1}{2+\frac{8(1-\gamma)}{\gamma}\rho^2}-b\right]x+c_2\left[\frac{\left(\frac{2(1-\gamma)}{\gamma}\rho^2-1\right)^2a^2c_2}{2+\frac{8(1-\gamma)}{\gamma}\rho^2}+\frac{a^2}{2}-bl\right]\frac{1}{x}\\
&+\frac{\frac{4(1-\gamma)}{\gamma}|\lambda\rho a|\left(1+\frac{1-\gamma}{\gamma}\rho^2\right)-\frac{2(1-\gamma)^2}{\gamma}a|\rho|^3C_p-2(1-\gamma)a|\rho|C_p}{1+\frac{4(1-\gamma)}{\gamma}\rho^2}\Big|c_1\sqrt{x}-\frac{c_2}{\sqrt{x}}\Big|.
\end{align*}
Due to $bl>\frac{a^2}{2}$, then sufficiently small positive constants $c_1$ and $c_2$ lead to $\mathfrak{F}[\phi](x)\searrow-\infty$ as $x\searrow0$ or $x\nearrow+\infty$. Therefore $\mathfrak{F}[\phi]$ is bounded from above on $(0,+\infty)$.

Step 3. Verification of Assumption \ref{rnm}.
According to Remark \ref{degen}, we only need to prove that for sufficiently small $\epsilon>0$,
\begin{align*}
&\mathbf{E}^0\left[\operatorname{exp}\left((1+\epsilon)(\psi-1)\int_0^Tr_s^+ds\right)\right]\\
\leq&\mathbf{E}^0\left[\operatorname{exp}\left((1+\epsilon)(\psi-1)\left(r_0^{+}T+r_1\int_0^TX_sds\right)\right)\right]<+\infty
\end{align*}
It remains to prove the integrability of $\operatorname{exp}\left((1+\epsilon)(\psi-1)r_1\int_0^TX_sds\right)$ under $\mathbb{P}_0$ . Indeed, using H\"older inequality once again, for $1<p',q'<+\infty$ with $\frac{1}{p'}+\frac{1}{q'}=1$,
\begin{align}\label{eh1}
&\mathbf{E}^0\left[\operatorname{exp}\left((1+\epsilon)(\psi-1)r_1\int_0^TX_sds\right)\right]\notag\\
=&\mathbf{E}\left[\mathscr{E}\left(\int-\lambda dW_s^\rho\right)_T\operatorname{exp}\left((1+\epsilon)(\psi-1)r_1\int_0^TX_sds\right)\right]\notag\\
\leq&\mathbf{E}\left[\mathscr{E}\left(\int-\lambda dW_s^\rho\right)_T^{p'}\right]^{\frac{1}{p'}}\mathbf{E}\left[\operatorname{exp}\left((1+\epsilon)q'(\psi-1)r_1\int_0^TX_sds\right)\right]^{\frac{1}{q'}}.
\end{align}
Following the above inequalities, one can similarly show that $\mathscr{E}\left(\int-\lambda dW_s^\rho\right)^{p'}$ is integrable. For the second expectation in (\ref{eh1}), we can choose $p'$,$q'$ sufficiently close to 1 such that according to Lemma \ref{LLT}, if
$$(\psi-1)r_1<\frac{b^2}{2a^2},$$
the second term is finite. This is exactly Condition (\romannumeral3) in Proposition \ref{exp1}.
$\hfill\Box$

\section{Conclusions}

The paper investigates the optimal consumption-investment problem for an investor with Epstein-Zin utility under general constraints in an incomplete market.  Closed constraints are imposed on strategies. Our method is based on BSDE techniques coming from \cite{HIM05} and  \cite{X17}.   The optimal consumption and investment strategies are characterized via a quadratic BSDE.   Comparing with \cite{X17},  the BSDE derived by the martingale optimal principle is more complicated,  while the upper boundedness of the solution still holds under appropriate assumptions.   In our situation,  the candidate optimal strategy cannot be explicitly expressed by $Z$.   To obtain the verification theorem,  we need to carry out more sophisticated estimations and delicately choose suitable Lyapunov function to overcome the difficulties.
Several examples  and  numerical simulations for the optimal strategies are  illustrated and discussed.

\section*{Acknowledgements}
Part of the work was completed by Dr. Tian during his visit to School of Mathematics, Shandong University. The warm hospitality of Shandong University is gratefully acknowledged. The authors would like to thank Prof. Shige Peng, the associate editor and two anonymous referees for their valuable comments and suggestions which led to a much improved
version of the paper.   This paper is supported by the National Natural Science Foundation of China (No. 12171471).

\end{document}